\documentclass[%
 reprint,
nofootinbib,
 amsmath,amssymb,
 aps,
]{revtex4-1}

\usepackage{graphicx}
\usepackage[dvipsnames]{xcolor}
\usepackage{dcolumn}
\usepackage{bm}
\usepackage[colorlinks]{hyperref}

\usepackage{xspace}
\usepackage{tikz}
\usepackage{adjustbox}

\newcommand{\Verne}{\textsc{Verne}\xspace}
\newcommand{\VerneII}{\textsc{Verne2}\xspace}

\newcommand{\DaMaSCUS}{\textsc{DaMaSCUS}\xspace}

\newcommand{\trimtikz}[1]{\trimbox{0.2cm .1cm .25cm 0.1cm}{#1}}

\usepackage{mathrsfs}
\usepackage{mathtools}

\newcommand\myshade{80}
\colorlet{mylinkcolor}{ForestGreen}
\colorlet{mycitecolor}{Red}
\colorlet{myurlcolor}{violet}

\hypersetup{
  linkcolor  = mylinkcolor!\myshade!black,
  citecolor  = mycitecolor!\myshade!black,
  urlcolor   = myurlcolor!\myshade!black,
  colorlinks = true
}

\begin{document}

\preprint{APS/123-QED}

\newcommand{\IFCA}{Instituto de F\'isica de Cantabria (IFCA, UC-CSIC), Av.~de
Los Castros s/n, 39005 Santander, Spain}

\title{A Fast Earth-scattering Formalism for Light Dark Matter with Dark Photon Mediators}

\author{Agust\'in Lantero-Barreda}
\email{lantero@ifca.unican.es}

\author{Carlos Centeno-Lorca}
\email{centeno@ifca.unican.es}

\author{Bradley J. Kavanagh}
\email{kavanagh@ifca.unican.es}

\author{N\'{u}ria Castello-Mor}
\email{castello@ifca.unican.es}

\affiliation{\IFCA}

\date{\today}

\begin{abstract}
While Dark Matter (DM) is typically assumed to interact only very weakly with the particles of the Standard Model, many direct detection experiments are currently exploring regions of parameter space where DM can have a large scattering cross section. In this scenario, DM may scatter in the atmosphere and Earth before reaching the detector, leading to a distortion of the DM flux and a daily modulation of the signal rate as the detector is shielded by more or less of the Earth at different times of day. This modulation is a distinctive signature of strongly-interacting DM and provides a powerful method of discriminating against time-independent backgrounds. However, the calculation of these Earth-scattering effects by Monte Carlo methods is computationally intensive, inhibiting a systematic exploration of the DM parameter space. Here, we present a semi-analytic formalism for calculating Earth-scattering effects, for models of MeV-mass DM which interacts via a dark photon mediator, and release the associated code \VerneII. This formalism assumes that DM travels along straight-line trajectories until it scatters and is reflected back along its incoming path, allowing us to take into account the affects of both attenuation and reflection in the Earth. We compare this formalism with the results of full Monte Carlo simulations for cross sections within reach of current and future DM-electron scattering searches. We find that \VerneII is accurate to better than 10-30\%, making it suitable for performing signal modeling in the search for daily modulation, while reducing the computational cost by a factor of $\sim10^4$ compared to full Monte Carlo simulations. 
\end{abstract}

\maketitle


\section{Introduction}

Uncovering the identity of Dark Matter (DM) is one of the greatest challenges in modern cosmology and particle physics~\cite{Bertone:2004pz,Bertone:2016nfn}. Direct detection experiments search for the scattering of DM particles from the halo of the Milky Way with nuclei or electrons in a target material~\cite{Drukier:1983gj,Goodman:1984dc,Drukier:1986tm}. Technological advances have pushed many direct detection experiments to lower and lower energy thresholds, opening the window to explore DM in the sub-GeV mass range~\cite{Billard:2021uyg}. While DM is typically assumed to have a very small scattering cross section with Standard Model particles, models with large cross sections remain viable (see e.g.~Refs.~\cite{1990PhRvD..41.3594S,Kavanagh:2017cru,Emken:2019hgy,Digman:2019wdm,Elor:2021swj}). This is especially true for light DM, where the expected signals are close to the energy thresholds of the detector and such large cross sections have not yet been excluded because of this reduced sensitivity. In these regions of the parameter space, the DM may scatter in the atmosphere and Earth before reaching the detector, distorting the incident DM flux and therefore the expected signal~\cite{Gould:1988eq, Collar:1993ss, Hasenbalg:1997hs}.

\begin{figure}[tb]
    \centering
    \includegraphics[width=0.8\linewidth]{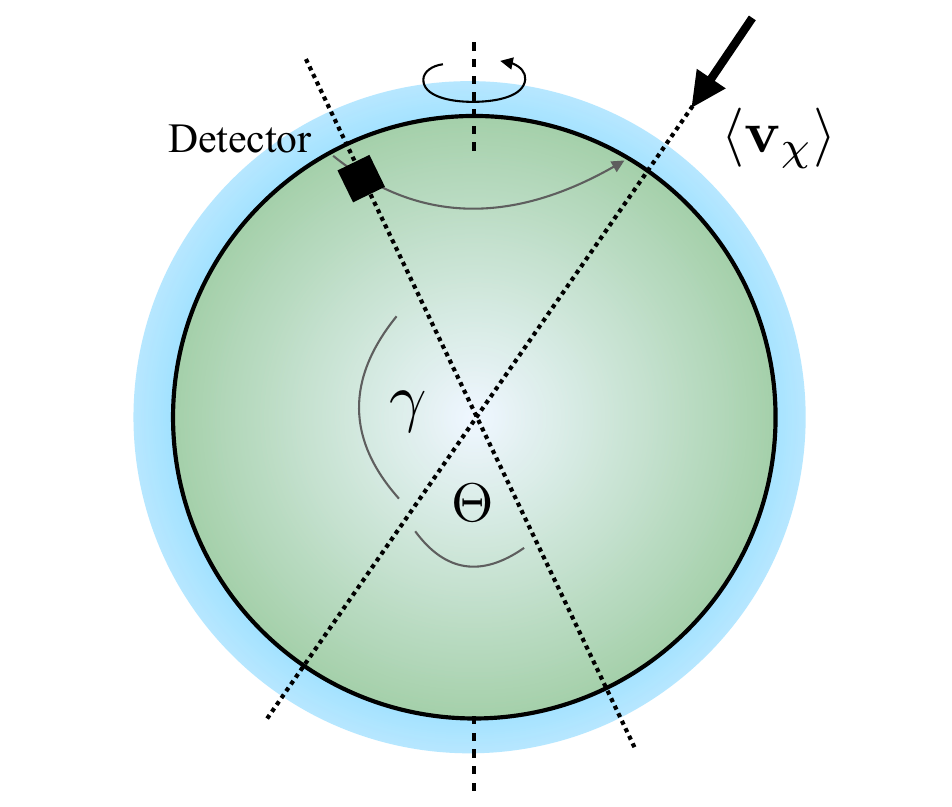}
    \caption{\textbf{Geometry of Earth-scattering Effects.} The mean DM velocity is fixed by $\langle \mathbf{v}_\chi \rangle = -\mathbf{v}_E$, where $\mathbf{v}_E$ is the velocity of the Earth in the rest-frame of the Milky Way. The distortion to the DM velocity distribution due to Earth-scattering depends on the isodetection angle $\gamma$ between the mean DM velocity and the local zenith of the detector (an alternative definition for the isodetection angle is $\Theta = 180^\circ - \gamma$~\cite{Emken:2017qmp}). As the Earth rotates, the detector traces different values of $\gamma$, leading to a daily modulation of the DM signal rate.}
    \label{fig:isodetection}
\end{figure}

A key signature of strongly-interacting DM is a characteristic diurnal modulation in the signal rate. As illustrated in Fig.~\ref{fig:isodetection}, the DM flux arriving at Earth has a preferential direction, set by the velocity of the Earth through the Milky Way Halo: $\langle \mathbf{v}_\chi \rangle = -\mathbf{v}_E$. The Earth-induced attenuation and reflection of DM particles alters the flux and velocity distribution of particles arriving at a detector. Owing to the azimuthal symmetry about $\langle \mathbf{v}_\chi \rangle$, these `Earth-scattering' effects depend only on the isodetection angle $\gamma$, defined as the angle between the mean DM flux and the local zenith at the detector. Over the course of a sidereal day, a detector will trace out different values of $\gamma$ as the Earth rotates, leading to a daily modulation of the expected DM signal rate~\cite{Kouvaris:2014lpa}. Backgrounds are not expected to be periodic on timescales of days, meaning that this daily modulation offers a powerful method of background discrimination, especially in the low-mass regime where background modeling can be challenging~\cite{Fuss:2022fxe}. However, taking advantage of this time-varying signal requires fast and accurate modeling of the relevant Earth-scattering effects.

The first efforts to estimate Earth-scattering effects in DM signals followed the energy loss of DM particles due to scattering along straight-line trajectories~\cite{1990PhRvD..41.3594S}. This approach has subsequently been refined and applied to estimate the sensitivity of direct detection experiments~\cite{Albuquerque:2003ei,Zaharijas:2004jv,Kouvaris:2014lpa,Kouvaris:2015laa,Davis:2017noy,Hooper:2018bfw,Cappiello:2023hza}. The assumption of straight-line trajectories is valid for super-heavy DM, where deflection is expected to be minimal and the assumption of continuous energy-loss may be justified~\cite{Kavanagh:2017cru,Bramante:2018tos,Bramante:2018qbc}. In the more general case, alternative approaches may be required, such as treating Earth-scattering as a diffusion-like process~\cite{Mahdawi:2017cxz, Mahdawi:2018euy}. Ultimately, Monte Carlo approaches, tracking individual DM trajectories and scattering events, provide the most complete description of these effects. Early Monte Carlo simulations focused on DM particles with moderate scattering cross sections and GeV-TeV masses~\cite{Gould:1988eq, Collar:1993ss, Hasenbalg:1997hs}. More recently, the public code \DaMaSCUS ~\cite{Emken:2017qmp,DAMASCUScode} has extended the regime of validity of Monte Carlo simulations to a wider range of masses and cross sections, allowing for scattering off both nuclei and electron~\cite{Emken:2017qmp, Emken:2018run, Emken:2019hgy}, and has been validated by semi-analytic calculations in the single-scatter regime~\cite{Kavanagh:2016pyr}. However, the accuracy and flexibility of full Monte Carlo simulations come at considerable computational cost.

In this paper, we present a novel semi-analytic formalism for calculating Earth-scattering effects for $\mathcal{O}(\mathrm{MeV})$ mass DM interacting  with the Standard Model via a dark photon mediator, a key target for direct searches~\cite{Battaglieri:2017aum,Knapen:2017xzo}. At the MeV-scale, kinematics favors searches for DM-electron scattering, as such light DM particles do not have sufficient energy to excite detectable nuclear recoils~\cite{Essig:2011nj}. However, the dominant Earth-scattering effects in such models typically come from DM-nucleus scattering~\cite{Emken:2019tni}. In addition, in this low-mass regime, the DM particles may be deflected through large angles by scattering off nuclei, meaning that the standard analytical approaches to Earth-scattering are not suitable. The formalism we present accounts for the attenuation and reflection of light DM particles which scatter up to twice before reaching the detector, applicable for the moderate DM scattering cross sections within reach of current detectors. 

The formalism and associated code \VerneII~\cite{Verne2} allow for a substantially faster evaluation of the daily modulation signal compared to full Monte Carlo methods.\footnote{\VerneII is publicly available at \url{https://github.com/bradkav/verne} and is an update and extension of the \Verne code originally presented in~\cite{Kavanagh:2017cru}.} For strongly-interacting DM, the signal spectrum must be re-computed for different cross sections, unlike in standard searches, in which the cross section only provides an overall rescaling of the rate. Fast signal evaluation is therefore crucial for a comprehensive exploration of the DM parameter space (see e.g.~Ref.~\cite{Bertou:2025adb}). Early versions of \VerneII have already been used by the DAMIC-M~\cite{DAMIC-M:2023hgj} and SENSEI~\cite{SENSEI:2025qvp} collaborations to derive world-leading limits on the daily modulation due to Dark Matter scattering in the mass range 0.5 - 2.7 MeV.

In Sec. \ref{sec:Scattering} we describe the model of DM interacting via dark photon mediators and provide the relevant scattering cross sections. In Sec.~\ref{sec:attenuation} we detail the new semi-analytical approach for the attenuation and reflection of DM in its scattering through the Earth. In Sec.~\ref{sec:results}, we present the velocity distributions and DM-electron scattering rates obtained using \VerneII for a range of benchmark parameter points. We compare with results from \DaMaSCUS, providing the first cross-validation of the code for scattering via dark photon mediators. We quantify the level of discrepancy between the signal rates obtained using the two codes and find that \VerneII typically produces results at the level of 10-30\% accuracy, at a computation cost $\sim 10^4$ times smaller than the full Monte Carlo approach.

\section{Dark Photon-mediated Dark Matter interactions} 
\label{sec:Scattering}

We focus on models of light DM which couples to a Dark Photon $A^\prime$~\cite{GALISON1984279,HOLDOM1986196,Fabbrichesi:2020wbt}. Kinetic mixing of the Dark Photon with the Standard Model photon induces an interaction between DM and the charged particles of the SM.\footnote{Strictly speaking, we should consider the mixing of the Dark Photon with the gauge boson associated with the $U_Y(1)$ hypercharge symmetry of the SM, but for our purposes, we do not miss any interesting phenomenology by considering the mixing direction with the visible photon.} The resulting differential cross-section for scattering off free electrons and protons is:
\begin{equation}
    \frac{\mathrm{d}\sigma_{e, p}}{\mathrm{d}q^2} = \frac{4 \pi \alpha \alpha_D \epsilon^2}{\left(q^2 + m_{A^\prime}^2\right)^2} \frac{1}{v^2}\,.
\end{equation}
Here, $q$ is the momentum transfer; $v$ is the DM velocity; $\alpha$ and $\alpha_D$ are the SM and dark fine-structure constants respectively; and $\epsilon$ is the kinetic mixing parameter between the SM and dark photons. It is also convenient to define the reference cross section:
\begin{equation}
    \bar{\sigma}_{e} = \frac{16 \pi \alpha \alpha_D \epsilon^2 \mu_{\chi e}^2}{\left(q_\mathrm{ref}^2 + m_{A^\prime}^2\right)^2}\,,
\end{equation}
and similarly define $\bar{\sigma}_p$ for scattering off protons. The DM-electron reduced mass is $\mu_{\chi e} = m_\chi m_e/(m_\chi + m_e)$ and we fix the reference momentum transfer to be $q_\mathrm{ref} = \alpha m_e$~\cite{Essig:2011nj}. 
With this, we can re-write the differential cross section as
\begin{equation}
 \frac{\mathrm{d}\sigma_{e}}{\mathrm{d}q^2} = \frac{\bar{\sigma}_e}{4 \mu_{\chi e}^2 v^2} F_\mathrm{DM}^2(q)\,,
\end{equation}
and similarly for protons, factorizing out the momentum dependence in terms of a DM form factor:
\begin{equation}
    F_\mathrm{DM}(q) = \frac{q_\mathrm{ref}^2 + m_{A^\prime}^2}{q^2 + m_{A^\prime}^2}\,.
\end{equation}
From these expressions it is now clear that the reference cross sections $\bar{\sigma}_{e, p}$ would be the total cross sections for scattering off free electrons and protons, in the absence of momentum dependence ($F_\mathrm{DM} \rightarrow 1$).

These DM-electron and DM-proton reference cross sections are related by:
\begin{align}
    \bar{\sigma}_p = \bar{\sigma}_e \left( \frac{\mu_{\chi p}}{\mu_{\chi e}}\right)^2 = \bar{\sigma}_e\left(\frac{m_p (m_\chi + m_e)}{m_e ( m_\chi + m_p)}\right)^2\,.
\end{align}
For DM in the MeV mass-range, the DM-proton cross section is typically larger than the DM-electron cross section (by a factor of order 10 or more). In fact, the DM-nucleus cross section is also coherently enhanced by a factor of $Z^2$, where $Z$ is the nuclear charge, making it larger still. Direct detection experiments typically search for DM in this mass range via DM-electron scattering. However, Earth scattering effects are typically dominated by DM-nucleus scattering due to the larger cross section~\cite{Lee:2015qva,Emken:2017erx,Emken:2019tni}.

In order to compute the DM-nucleus scattering cross-section, we must also take into account the effects of screening by atomic electrons; at large distances (low momentum transfer) the nucleus appears to be effectively neutral~\cite{Kaplinghat:2013yxa,Emken:2019tni}. The relevant length-scale for this screening is set by the Thomas-Fermi radius, given by $a  \approx 0.89 Z^{-1/3} a_0$, where $a_0$ is the Bohr radius~\cite{doi:10.1142/9789812795755_0003}. Taking into account this atomic screening and integrating over momenta, the total DM-nucleus cross section can be expressed in terms of the reference cross-section as~\cite{Emken:2019tni}:
\begin{align}\label{eq:sigma_N}
\begin{split}
    \sigma_N (v) &= \bar{\sigma}_p \left( \frac{\mu_{\chi N}}{\mu_{\chi p}} \right)^2 Z^2\times \\
    &\begin{dcases}
    1 + \frac{1}{1 + x} - \frac{2}{x}\log\left(1 + x\right) \,,&\textrm{(HM)}\\
    \frac{\left(a \,q_\mathrm{ref}\right)^4}{1 + x}\,. & \textrm{(ULM)}
    \end{dcases}
\end{split}
\end{align}
 The parameter $x$ is given by $x \equiv a^2 q_\mathrm{max}^2$, with the maximum momentum transfer $q_\mathrm{max} = 2 \mu_{\chi N} v$. Here, we distinguish between the two limiting cases for the mediator mass. For the heavy mediator (HM) scenario, $q^2 \ll  m_{A^\prime}^2$, the dark photon effectively mediates a contact interaction with $F_\mathrm{DM} = 1$. In the ultra-light mediator (ULM) scenario, $q^2 \gg  m_{A^\prime}^2$, the dark photon mediator is effectively massless, leading to a form factor $F_\mathrm{DM} \propto 1/q^2$.

\begin{figure}[tb]
    \centering
    \includegraphics[width=0.98\linewidth]{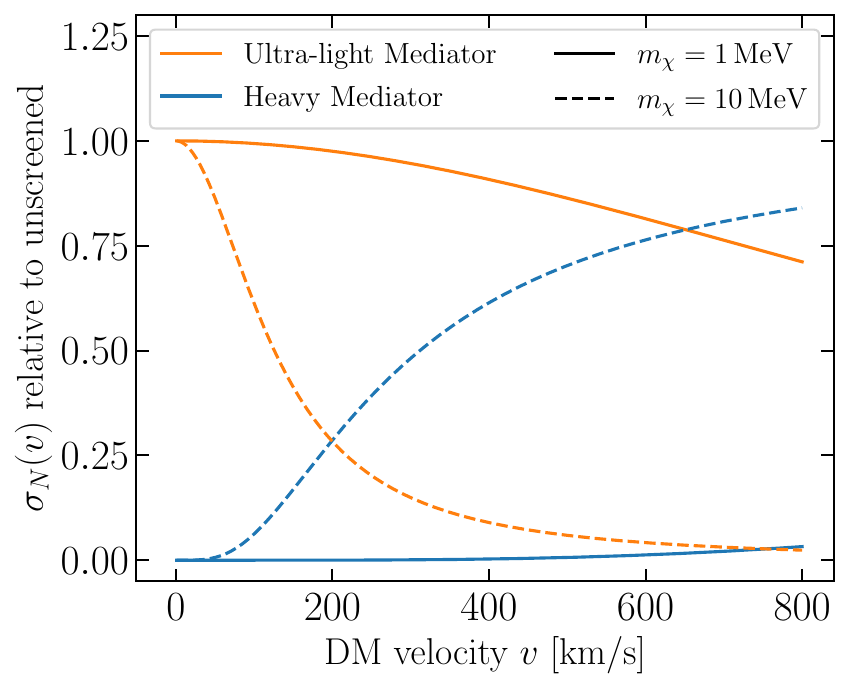}
    \caption{\textbf{Total DM-nucleus scattering cross section for dark photon mediated interactions, relative to the unscreened case.} We assume scattering off Oxygen nuclei. In the ultra-light mediator case (orange), the total cross section saturates to $(a\,q_\mathrm{ref})^{4}$ times the unscreened cross section at low velocity (we have factored this numerical value out for the ultra-light mediator, leading to saturation at 1). In the heavy mediator case (blue), the total cross section is zero at low velocity, saturating to the unscreened value at large velocities.}
    \label{fig:cross_section}
\end{figure}

In Fig.~\ref{fig:cross_section}, we illustrate how the DM-nucleus scattering cross section varies with velocity. We normalize the total DM-nucleus cross section to its value in the absence of screening\footnote{In the ultra-light mediator case, we also rescale by a factor of $(a\,q_\mathrm{ref})^{-4}$.}. Solid lines show the results for a DM particle with mass $m_\chi = 1\,\mathrm{MeV}$ scattering off Oxygen nuclei (which are the most abundant in the Earth's crust). In the ultra-light mediator case, the cross section is suppressed with increasing DM velocity. Instead, for the heavy mediator, the cross section is zero at low velocity, increasing and ultimately saturating at the unscreened value for sufficiently large velocities.  
The onset of screening effects happens at a lower velocity as we increase the DM mass (dashed lines for $m_\chi = 10\,\mathrm{MeV}$). This is because, at fixed $v$, an increase in the DM mass leads to an increase in the maximum momentum transfer $q_\mathrm{max}$ and therefore also in the screening parameter $x$.

\section{Earth scattering formalism}
\label{sec:attenuation}

We now describe the semi-analytic formalism which was developed for the \VerneII code, in order to allow for the fast evaluation of the DM velocity distribution at the detector, taking into account Earth-scattering effects. As described above, we only take into account DM-nucleus scattering during the propagation of particles through the atmosphere and Earth. In addition, our focus will be on light (i.e. MeV-scale) DM particles with moderate scattering cross sections (for which the number of expected scattering events during the DM's transit through the Earth is of the order of a few).  

With this in mind, we make a number of simplifying assumptions:
\begin{itemize}
    \item We assume that the DM particles do not lose energy when they scatter.
    \item We account for \textit{at most} 2 scattering events before the DM particle reaches the detector. 
    \item When a DM particle scatters, it either continues travelling in the same direction in a straight line (with probably $p_\mathrm{forward}$) or it is reflected back along its incoming direction (with probability $p_\mathrm{back} = 1 - p_\mathrm{forward}$). 
\end{itemize}
The first assumption is justified  by the fact that when light DM scatters off a much heavier nucleus $m_\chi \ll m_N$, it loses a fraction $\Delta E/E \sim m_\chi/m_N \ll 1$ of its initial energy~\cite{Kavanagh:2016pyr}. The second assumptions limits the validity of the approach to moderate cross-sections, though as we will see in Sec.~\ref{sec:results}, this assumption remains valid over much of the parameter space accessible to current and near-future direct detection experiments.

The third assumption allows us to ignore the full three-dimensional scattering of the DM particles (as would be tracked in a Monte Carlo simulation), instead allowing us to focus on straight line trajectories. We next calculate the probability of this backward scattering.

\subsection{Deflection angles}

After scattering off a nuclear target in the Earth or atmosphere, the DM particle will be deflected from its initial trajectory by an angle $\alpha$. The probability distribution for $\alpha$ is given by:
\begin{equation}
    P(\mathrm{cos}\,\alpha)\propto \left|\frac{\mathrm{d}\sigma_N}{\mathrm{d}\mathrm{cos}\,\alpha}\right| \propto \frac{m_\chi^2 v^2}{m_N} \left|\frac{\mathrm{d}\sigma_N}{\mathrm{d}q^2}\right|\,,
\end{equation}
where we have taken advantage of the fact that for $m_\chi \ll m_N$, the momentum transfer is given by $q^2 = (2 m_\chi^2 v^2/m_N)(1 - \cos\alpha)$~\cite{Kavanagh:2016pyr}. Performing the change of variables explicitly, we obtain:
\begin{align}
    \label{eq:deflection}
    P(\cos\alpha) &= N_\mathrm{H} \frac{(1 - \cos\alpha)^2}{\left(1 + \frac{1}{2}x (1-\cos\alpha)\right)^2}\,, & \mathrm{(HM)}  \\
    P(\cos\alpha) &= N_\mathrm{UL} \frac{1}{\left(1 + \frac{1}{2}x (1-\cos\alpha)\right)^2}\,,  & \mathrm{(ULM)}
\end{align}
where the normalizing constants are:
\begin{align}
    N_\mathrm{H} &= \frac{x^3 (1 + x)}{8 x (2 + x) - 16 (1 + x) \log(1 + x) }\,,\\
    N_\mathrm{UL} &= \frac{1 + x}{2}\,.
\end{align}

\begin{figure}[tb]
    \centering
    \hspace{-0.2cm}\includegraphics[width=0.99\linewidth]{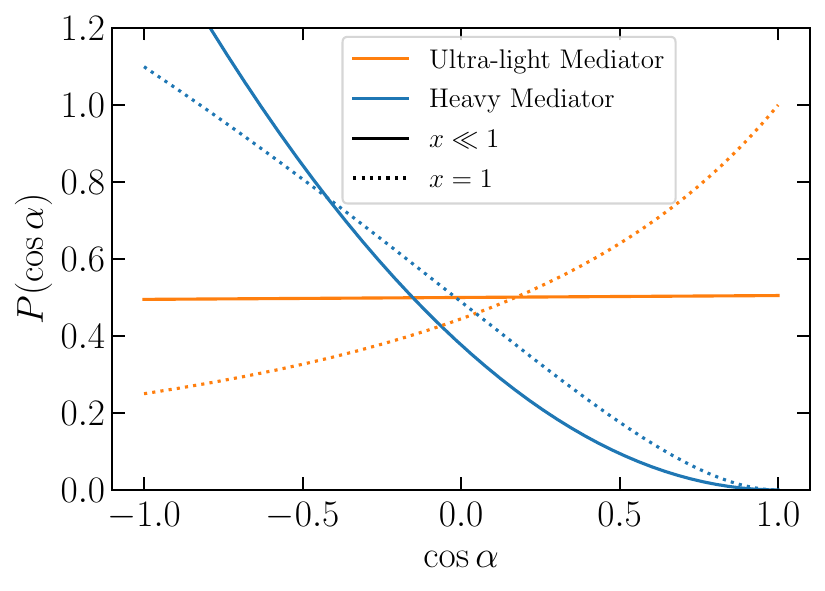}
    \caption{\textbf{Distribution of the deflection angle $\alpha$.} Note that $\cos\alpha = 1$ corresponds to forwards scattering while $\cos\alpha$ corresponds to backwards scattering.}
    \label{fig:Pcosalpha}
\end{figure}

In Fig.~\ref{fig:Pcosalpha}, we show the distribution of the scattering angle for $x = 1$ (dashed lines) and for $x\ll 1$ (solid lines). In this latter limit (valid for very light DM masses), the heavy mediator leads predominantly to backwards scattering, while the ultra-light mediator leads to an isotropic deflection distribution. In the rest of our analysis (and in the \VerneII code) we will consider the limit $x \rightarrow 0$, which is justified for light MeV DM.\footnote{A particle moving with the maximum velocity in the Earth frame $v  \sim 800$ km/s and a mass of $\sim 1\,\mathrm{MeV}$ will have a value of $x\sim0.4$.}

To simplify the propagation, we approximate any large-angle scatter ($\alpha > 90^\circ$) as a backward reflection (i.e.~the particle is reflected back along its original trajectory). As is clear from Fig.~\ref{fig:Pcosalpha}, this assumption is appropriate for the heavy mediator case, where backward scattering dominates, but less justified for the ultra-light mediator, for which the deflection is nearly isotropic.
In order to quantify the probability of backward scattering, we use the distributions in Eq.~\eqref{eq:deflection} to calculate:
\begin{equation}
    p_\mathrm{back} = \int_{-1}^0 P(\mathrm{cos}\alpha)\,\mathrm{d}\cos\alpha\,.
\end{equation}
This gives the fraction of scattering events in which the DM is deflected through more than $90^\circ$, into the backwards hemisphere. In the limit $x\rightarrow 0$, we obtain:
\begin{align}\label{eq:Phl}
    p_\mathrm{back}=
    \begin{cases}
    \frac{7}{8}\,, & (\mathrm{HM})\\
    \frac{1}{2}\,. & (\mathrm{ULM})
    \end{cases}
\end{align}

\subsection{Scattering probabilities}
\label{sec:probabilities}

\begin{figure*}[tb]
\centering
\includegraphics[width=.4\textwidth]{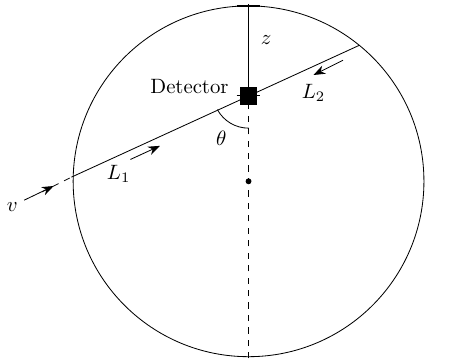}
\includegraphics[width=.41\textwidth]{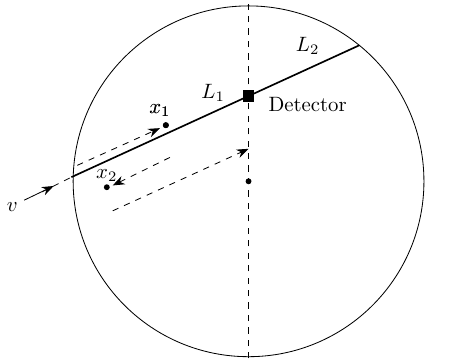}
\caption{\textbf{Scheme for attenuation and reflection of DM particles.} (a) DM particle path to the detector with 0-scatters and 1-scatter. The unscattered particle will traverse $L_1$ and reach the detector. Once it passes the detector, the particle can be reflected back again along the path $L_2$. (b) 2-scatter scheme of DM through the Earth. The dashed arrows represents the trajectory of the particle after backwards scatter at $x_1$ anywhere within the path $L_1$, followed by a backwards scatter at $x_2$ anywhere within the path $L_1$ before the detector. }
\label{fig:scheme_probs}
\end{figure*} 

We now turn to calculating the probability of arrival at the detector, taking into account Earth-scattering effects. The scheme of this calculation is illustrated in Fig.~\ref{fig:scheme_probs}. Under the assumption of straight-line trajectories, it is useful to define two distances: the distance from the entrance point of the Earth (the upper atmosphere) to the detector $L_1$ and the distance between the exit point and the detector $L_2$. These distances are given by:
\begin{align}
    L_1(\theta) &= r_\mathrm{det}\cos\theta + \sqrt{(R_\mathrm{E} + h_\mathrm{A})^2 - (r_\mathrm{det}\sin\theta)^2}\\
    L_2(\theta) &= 2(R_\mathrm{E} + h_\mathrm{A}) - L_1(\theta)\,,
\end{align}
where the radius of the Earth is $R_E \approx 6370\,\mathrm{km}$, the height of the atmosphere is $h_A \approx 80 \,\mathrm{km}$ and we define $r_\mathrm{det} = R_\mathrm{E} - z$ for a detector at a depth $z$ below the Earth's surface. The paths along $L_1$ and $L_2$ depend on the direction of the incoming DM trajectory. This is specified by the polar angle $\theta = \cos^{-1} \hat{\textbf{v}}\cdot\hat{\textbf{z}}$, where $\hat{\textbf{v}}$ is the unit vector parallel to the incoming DM velocity and $\hat{\textbf{z}}$ is the unit vector pointing towards the local zenith at the detector. With this, $\theta = 0^\circ$ corresponds to DM particles arriving from directly below the detector while $\theta = 180^\circ$ corresponds to particles arriving from directly above.

The flux arriving at the detector can then be decomposed into two components: the transmitted flux and the reflected flux. The transmitted flux consists of those particles which arrive at the detector without scattering over the distance $L_1$ (left panel of Fig.~\ref{fig:scheme_probs}) or those which arrive after having scattered twice over that distance (right panel of Fig.~\ref{fig:scheme_probs}).\footnote{Here, we ignore forward scattering events, as these do not affect the energy or trajectory of the particle. Hence, each scattering event reverses the direction of the incoming particle.} The reflected component consists of those particles which scatter after passing the detector, over the distance $L_2$, and which are then reflected back. As a function of these distances, we can calculate the probabilities of $0$-scatter ($P_0$), $1$-scatters ($P_1$) and $2$-scatters ($P_2$).

The probability that a particle traverses some straight-line path $C$ (of length $L$) without scattering backwards is given by:
\begin{equation}\label{eq:P0}
    P_{0}(L,v) = \mathrm{exp} \left[ -p_\mathrm{back}\sum_i^\mathrm{species} \sigma_i(v) \int_C n_i(\textbf{r}) \,\mathrm{d}l\right] \,.
\end{equation}
Here, $\sigma_i(v)$ is the total DM scattering cross section off a nuclear species $i$, as given in Eq.~\eqref{eq:sigma_N}, and $n_i(\mathbf{r})$ is the number density of that species.
It is convenient also to define the effective back-scattering probability over a distance $L$ as
\begin{equation}\label{eq:D_Lambda}
       p_\mathrm{eff}(L, v) = p_\mathrm{back}   \sum_i^\mathrm{species} \int_C  \frac{\mathrm{d}l}{\lambda_i(\textbf{r},v)} \,,
\end{equation}
where $\lambda_i(\textbf{r},v) = n_i(\textbf{r})\sigma_i(v)$ is the mean free path of the DM for scattering off nuclei of type $i$ in the Earth or atmosphere. 

The integral over the path length in Eq.~\eqref{eq:D_Lambda} is performed from the upper atmosphere, through the Earth, to the detector, along the trajectory defined by $\theta$. For the composition of the atmosphere and Earth we consider:
\begin{itemize}
    \item \textbf{Atmosphere}. Formed by Oxygen and Nitrogen and extending to a height of 80 km above the surface of the Earth. The density profiles are obtained from Ref.~\cite{Atmosphere}.
    \item \textbf{Earth}. We assume a model formed by 8 different elements: O, Si, Mg, Fe, Ca, Na, S, Al. The density profiles are obtained from Refs.~\cite{ELayer2,ELayer3} (previously tabulated in Ref.~\cite{ELayer1}).
\end{itemize}
This model for the density of elements in the atmosphere and Earth matches that used in the original \Verne code~\cite{Verne,Kavanagh:2017cru}.

In order to calculate the contribution to the flux from 2-scattering events, we first write down the probability of travelling a distance $x$ without scattering, before scattering once in the infinitesimal interval $\mathrm{d}x$:
\begin{equation}
\label{eq:P_scatter}
    P_\mathrm{scat}(x, v)\,\mathrm{d}x = e^{-p_\mathrm{eff}(x, v)}\,p_\mathrm{eff}(x, v)\,\mathrm{d}x\,.
\end{equation}
To calculate the probability of having 2-scatters, we will need to obtain the individual probability of scattering once at any point $x_1$, in the straight path followed by the particle within $L_1$. Then we will need to calculate the probability of scattering once again in the path from $x_1$ up to any other $x_2$, where the particle will be reflected back to the detector. Finally, we will need to calculate the probability of remaining unscattered from $x_2$ to the detector at $L_1$. The scheme of this process is shown in Fig~\ref{fig:scheme_probs}(b). The total probability is then obtained from
\begin{align}
\begin{split}
\label{eq:P2_general}
    P_2(L) &= \int_0^L dx_1 \int_{0}^{x_1}dx_2 \cdot \\
    &P_\mathrm{scat}(x_1)P_\mathrm{scat}(x_2 \leftarrow x_1)P_{0}(x_2\rightarrow L) \,,
    \end{split}
\end{align}
where we have dropped the explicit dependence on $v$ for notational clarity.
Now Eq.~\eqref{eq:P2_general} can be solved by using the appropriate distances in Eq.~\eqref{eq:P_scatter}. The result of the integral is:
\begin{equation}\label{eq:P2}
    P_2(L) =  \left(\frac{p_\mathrm{eff}(L)}{2} - \frac{1}{4}\right) e^{-p_\mathrm{eff}(L)}  +\frac{1}{4} e^{-3p_\mathrm{eff}(L)} \,,
\end{equation}
where $L$ here can be $L_1$ or $L_2$, as this expression is general for any given path.

In principle, to obtain the reflected component, we can perform a similar calculation, accounting for single scattering events which occur over the distance $L_2$ on the far side of the detector. However, in order to conserve probability, we fix the value of the single-scatter contribution based on the $0$- and $2$-scatter contributions:
\begin{equation}
    P_1(L) = 1 - P_0(L) - P_2(L)\,.
\end{equation}

\subsection{Velocity distribution}

The DM velocity distribution at the detector can be written as:
\begin{align}
    f(v, \theta, \phi,\gamma) = p(v,\theta)f_i(v,\theta, \phi, \gamma)\,,
\end{align}
where $f_i$ is the initial (free) velocity distribution and the correction to the flux $p(v,\theta)$ is given by:
\begin{align}
p (v,\theta) = p_\mathrm{trans}(v, \theta) + p_\mathrm{refl}(v, \theta)\,.
\end{align}
Here, the angle $\phi$ is the azimuthal angle in spherical coordinates, measured around the axis defined by $\theta$ (see Fig.~\ref{fig:scheme_probs}). The angle $\gamma$ is the isodetection angle, defined as the angle between the local zenith and the mean direction of the incoming DM flux (see Fig.~\ref{fig:isodetection}). 
As discussed in the previous section, the transmission probability is given by:
\begin{align}
    p_\mathrm{trans}(v, \theta) = P_0(L_1, v) + P_2(L_1, v)\,.
\end{align}
The reflection probability can then be written as:
\begin{align}
    p_\mathrm{refl}(v, \theta) &= \left[P_0(L_1, v) + P_2(L_1, v)\right] \times P_1(L_2, v)\,,
\end{align}
where the first term is square brackets corresponds to the probability of being transmitted to the detector and the term $P_1(L_2, v)$ is the probability of being reflected backwards after passing the detector.

\begin{figure}
    \centering
    \hspace{-0.5cm}
    \includegraphics[width=1.05\linewidth]{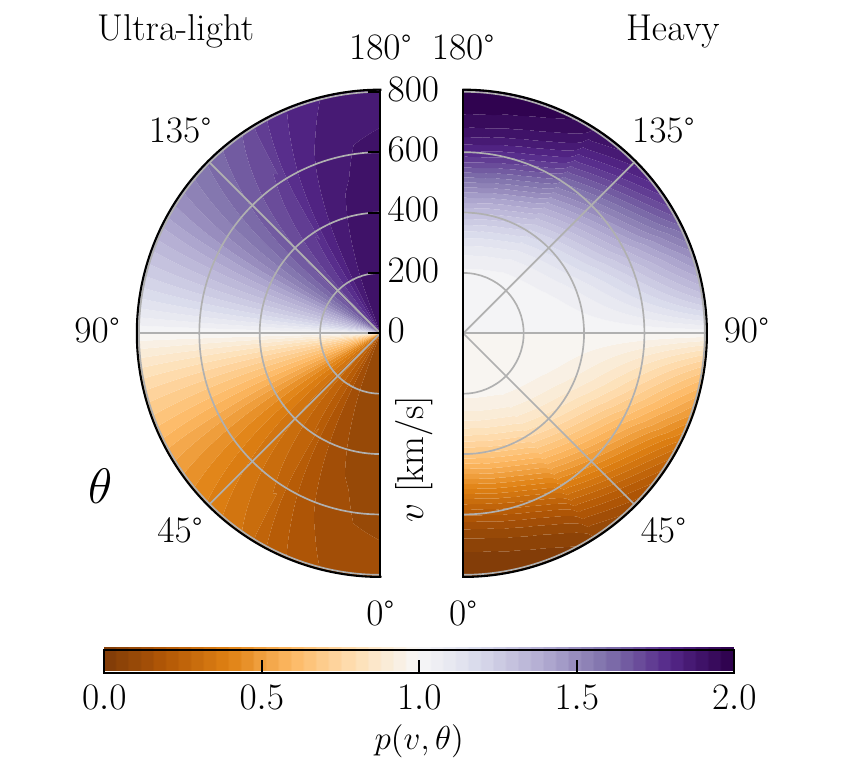}
    \caption{\textbf{Velocity distribution correction due to Earth scattering $p(v,\,\theta)$}, for a given DM velocity $v$ and incoming direction $\theta$. Particles arriving from below the detector (and crossing most of the Earth) have $\theta = 0^\circ$. For both the ultra-light mediator (left) and the heavy mediator (right), we fix the DM mass and cross section to be $m_\chi = 1\,\mathrm{MeV}$ and $\bar{\sigma}_p = 10^{-32}\,\mathrm{cm}^2$. }
    \label{fig:flux_correction}
\end{figure}

With these definitions, the correction to the velocity distribution due to Earth-scattering $p(v,\,\theta)$ is shown in Fig.~\ref{fig:flux_correction} for $m_\chi = 1\,\mathrm{MeV}$ and $\bar{\sigma}_p = 10^{-32}\,\mathrm{cm}^2$. For the heavy mediator, the DM-nucleus cross section grows with increasing DM velocity leading to almost complete depletion of the incoming DM flux for $v \gtrsim 700\,\mathrm{km}/\mathrm{s}$ and $\theta = 0^\circ$ (DM particles coming from below). By contrast, the cross section for the ultra-light mediator is increasingly suppressed at high velocity. Indeed, in that case the largest probability of attenuation ($\theta = 0^\circ$) and reflection ($\theta = 180^\circ$) occurs for velocities $v \lesssim 600\,\mathrm{km/s}$.

The full velocity distribution at the detector is then obtained by integrating over incoming DM angles:
\begin{equation}
    f(v, \gamma) = v^2 \int p(v,\theta)f_i(v,\theta, \phi, \gamma)\,\mathrm{d}\cos\theta\,\mathrm{d}\phi \,.
\end{equation}
In practice, we tabulate the integral over the angle $\phi$ as this does not depend on the DM properties, due to the azimuthal symmetry. We then perform the integral over $\theta$ numerically on a fine grid.  Note that with this definition, the velocity distribution arriving at the detector is no longer normalized to unity when integrating over velocities. We keep the local DM density fixed to $\rho_\chi = 0.3\,\mathrm{GeV}/\mathrm{cm}^2$~\cite{Read:2014qva,Green:2017odb} and absorb the variation in the density due to Earth-scattering in the overall normalization of $f(v, \gamma)$. A normalization greater than one corresponds to an enhancement in the total number density and a normalization less than one corresponds to a suppression of the number density with respect to the unscattered case. 

\section{Numerical comparisons}
\label{sec:results}

We now present detailed comparisons of the numerical velocity distributions obtained using \VerneII and with full Monte Carlo simulations using \textsc{DaMaSCUS}~\cite{DAMASCUScode}.

We focus on the region of parameter space at low mass ($\sim$MeV scale), which is currently being explored and where the current limits allow for a large enough cross section that Earth-scattering can be significant. For the ultra-light mediator, we consider DM-proton cross sections of $\bar{\sigma}_p = \{10^{-35},\,10^{-33},\,10^{-31}\}\,\mathrm{cm}^2$. For the heavy mediator, we also additionally consider $\bar{\sigma}_p = 10^{-29}\,\mathrm{cm}^2$. For both cases, we consider masses of $m_\chi = \{0.53,\, 1.0,\,2.7,\,10\} \,\mathrm{MeV}$. \footnote{The ratio of DM-electron to DM-proton reference cross sections is $\bar{\sigma}_e/\bar{\sigma}_p \approx \left\{0.24\,,0.11,\,0.025,\, 0.0024\right\}$ for DM masses of $m_\chi = \{0.53,\, 1.0,\,2.7,\,10\} \,\mathrm{MeV}$ respectively.} Below $m_\chi \approx 0.53 \,\mathrm{MeV}$ existing DM-electron searches rapidly lose sensitivity.

\paragraph{Velocity Parameters.} For the initial velocity distribution, we assume the Standard Halo Model: a Maxwell-Boltzmann distribution with a hard truncation at the escape velocity in the Milky Way rest frame~\cite{Lewin:1995rx,Green:2011bv}. We fix the local Galactic escape speed as $v_{\rm esc} =  544\,\text{km s}^{-1}$~\cite{Smith:2006ym,Piffl:2013mla}; the Local Standard of Rest to be $v_0 = 220\,\mathrm{km/s}$~\cite{Green:2017odb} and the Earth's velocity to be $v_E = 220.8\,\mathrm{km/s}$.\footnote{This value corresponds to the Earth's velocity (in the rest frame of the Milky Way) in November 2024, taking into account the motion of the Earth around the Sun~\cite{McCabe:2013kea}.}

\paragraph{DaMaSCUS Simulations.} Monte Carlos simulations were performed with the \texttt{chargescreening} branch\footnote{\url{https://github.com/temken/DaMaSCUS/tree/chargescreening}} of \DaMaSCUS version 1.1, which implements DM-nuclear scattering via either a heavy or ultra-light dark photon mediator. In order to obtain sufficient statistics in the high velocity tail, we perform 20 DaMaSCUS simulations at each parameter point ($m_\chi$, $\bar{\sigma}_p$). Each simulation uses 36 rings in isodetection angle and $10^6$ particles sampled in each ring. The velocity distribution is obtained by binning in velocity and, in each bin, calculating the median value of $f(v)$  across simulations. The \DaMaSCUS simulations typically took around 1800 CPU-hours per parameter point for the highest cross sections and around 360 CPU-hours per point for the smallest cross sections. In comparison, the \VerneII calculations take approximately 10 minutes per point.

\subsection{Velocity Distributions}

\begin{figure*}[tb]
    \centering
    \includegraphics[width=0.31\linewidth]{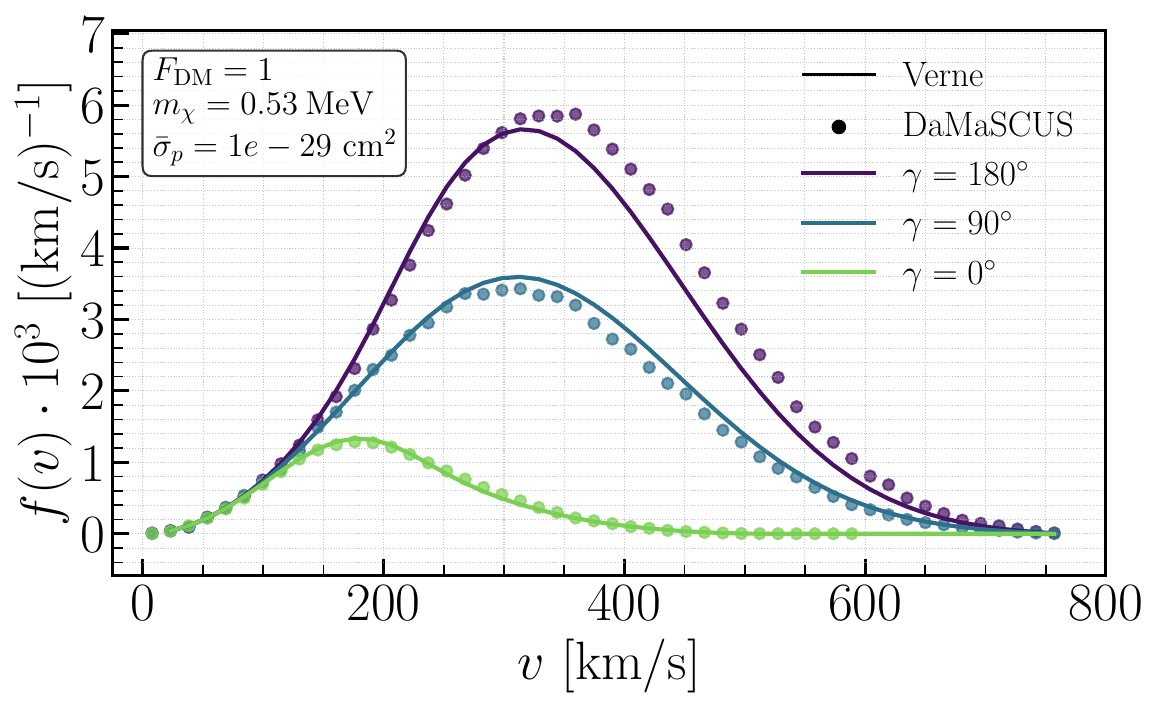}
    \includegraphics[width=0.31\linewidth]{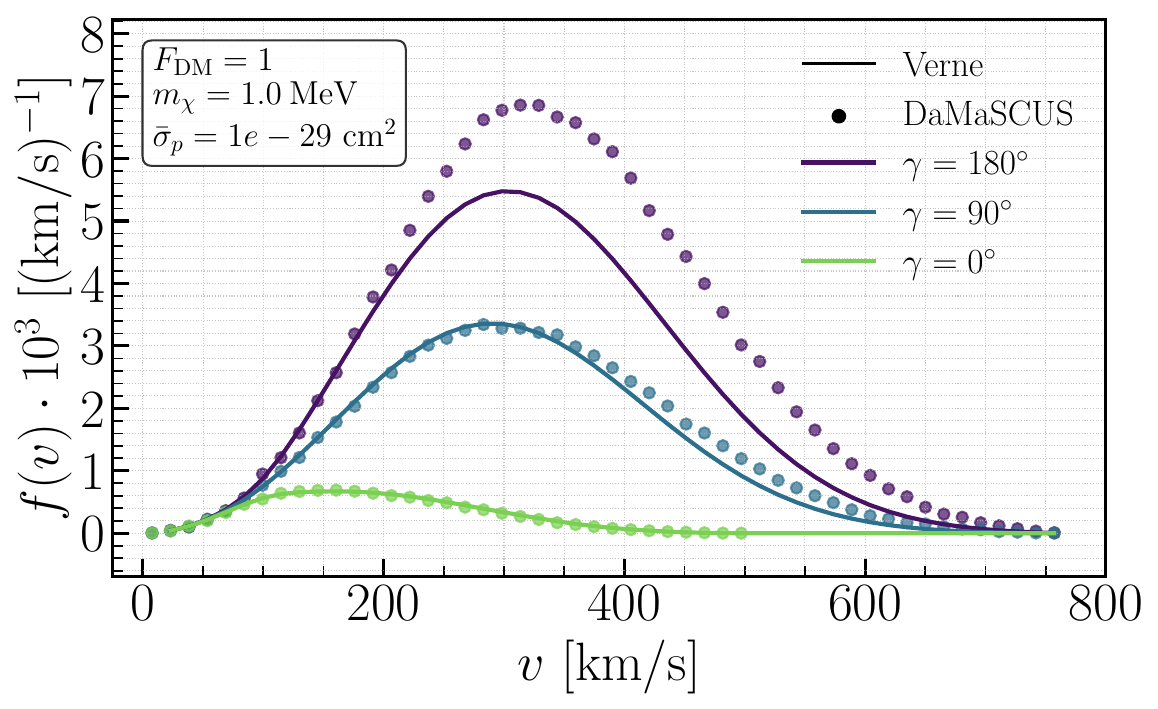}
    \includegraphics[width=0.31\linewidth]{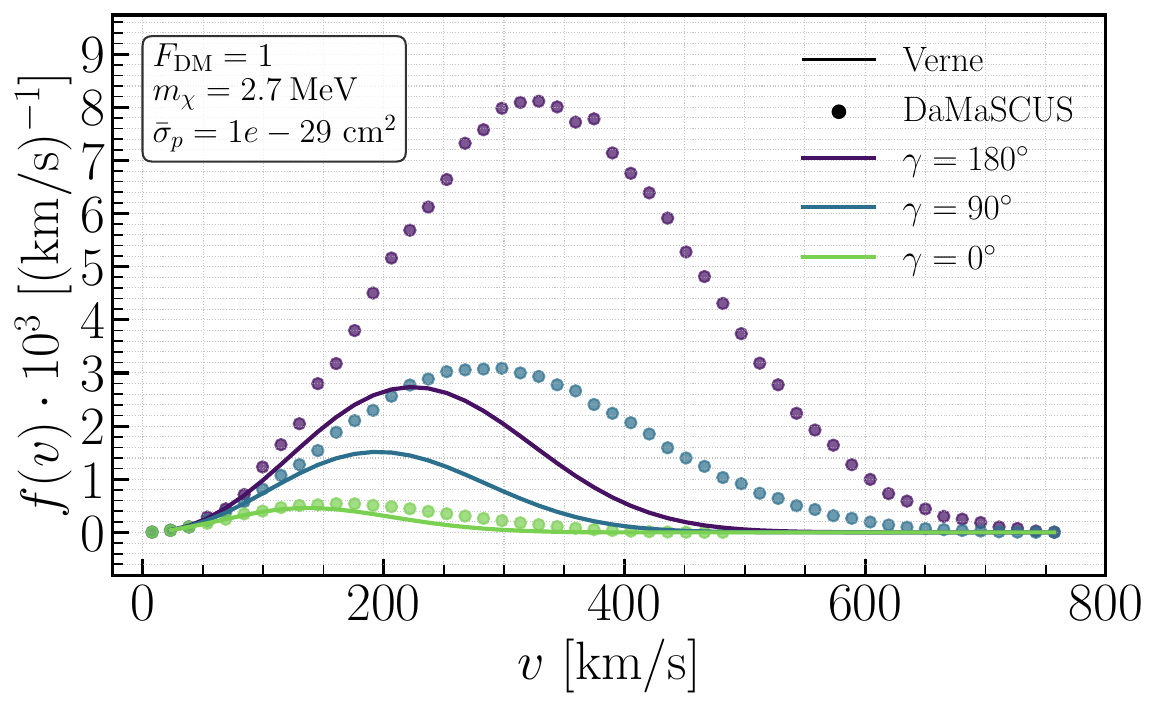}
    \includegraphics[width=0.31\linewidth]{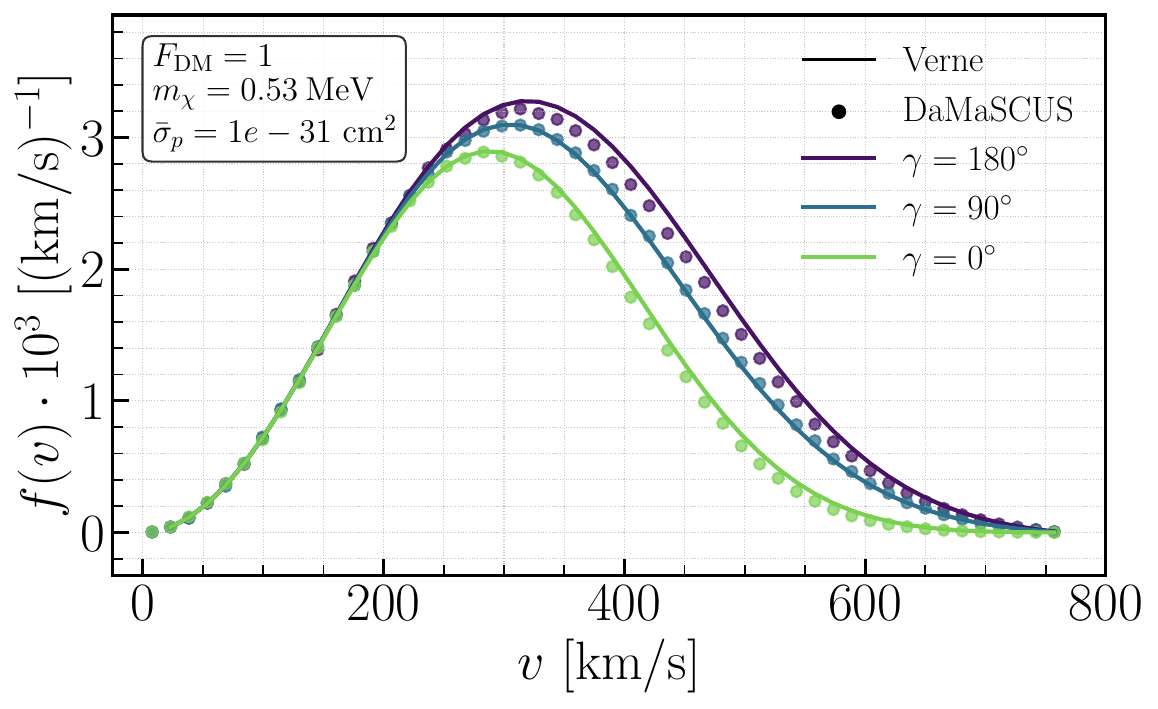}
    \includegraphics[width=0.31\linewidth]{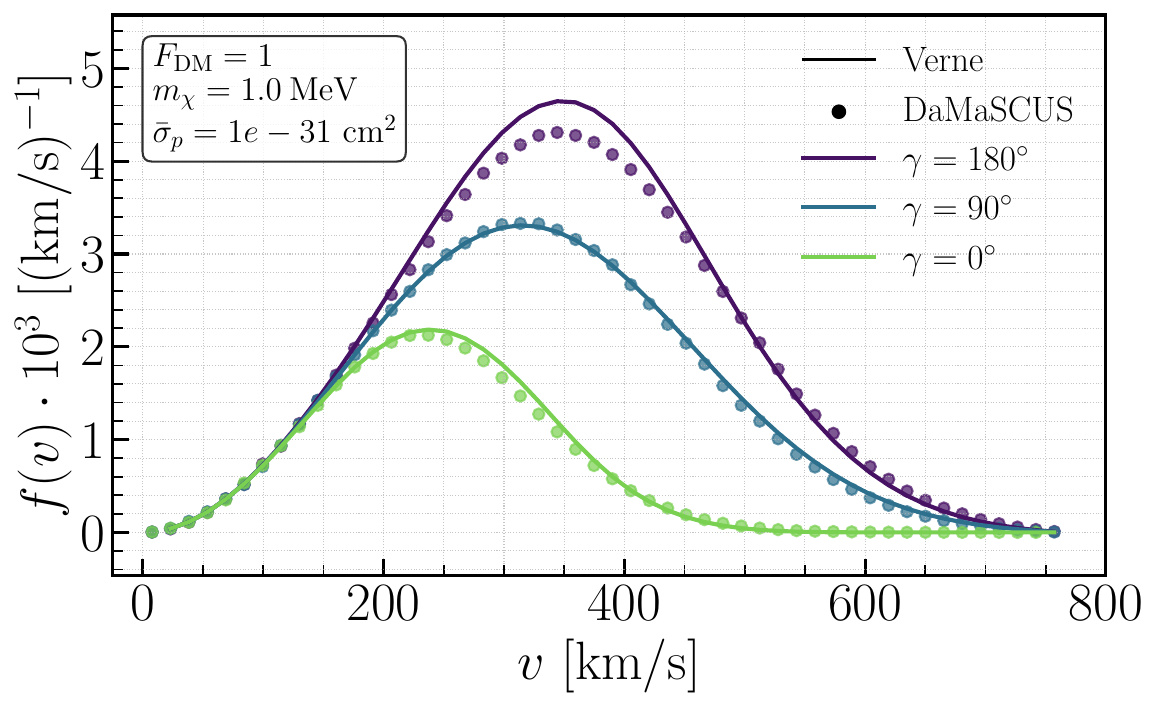}
    \includegraphics[width=0.31\linewidth]{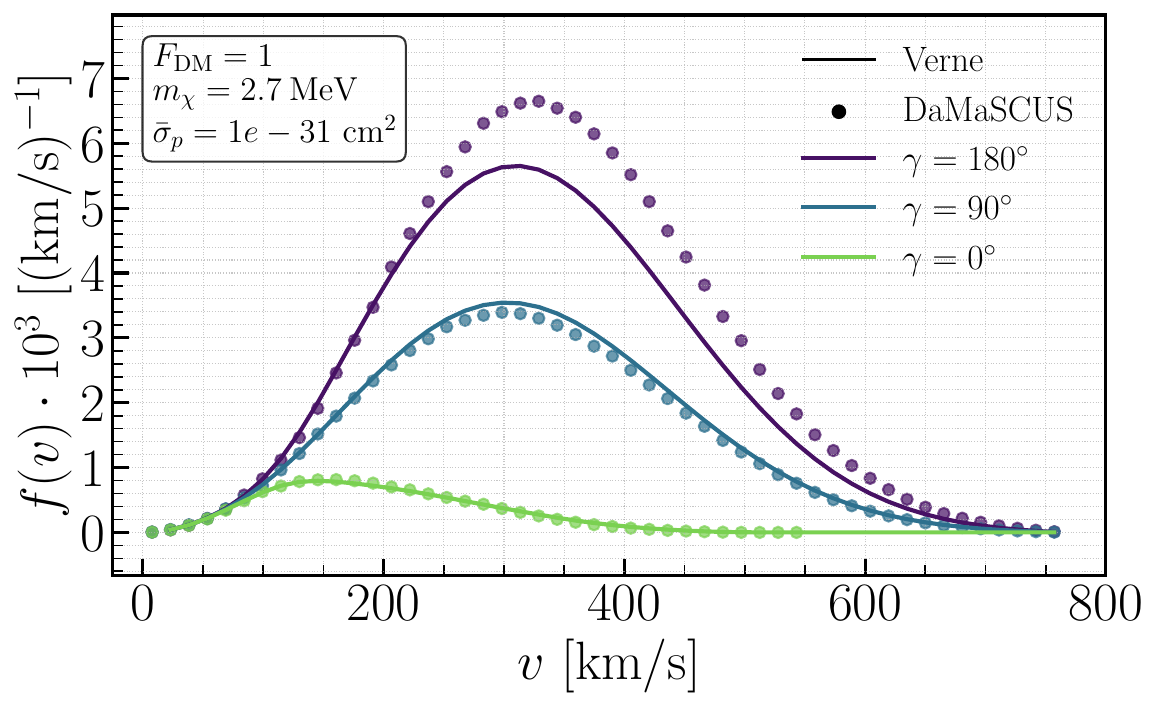}
    \includegraphics[width=0.31\linewidth]{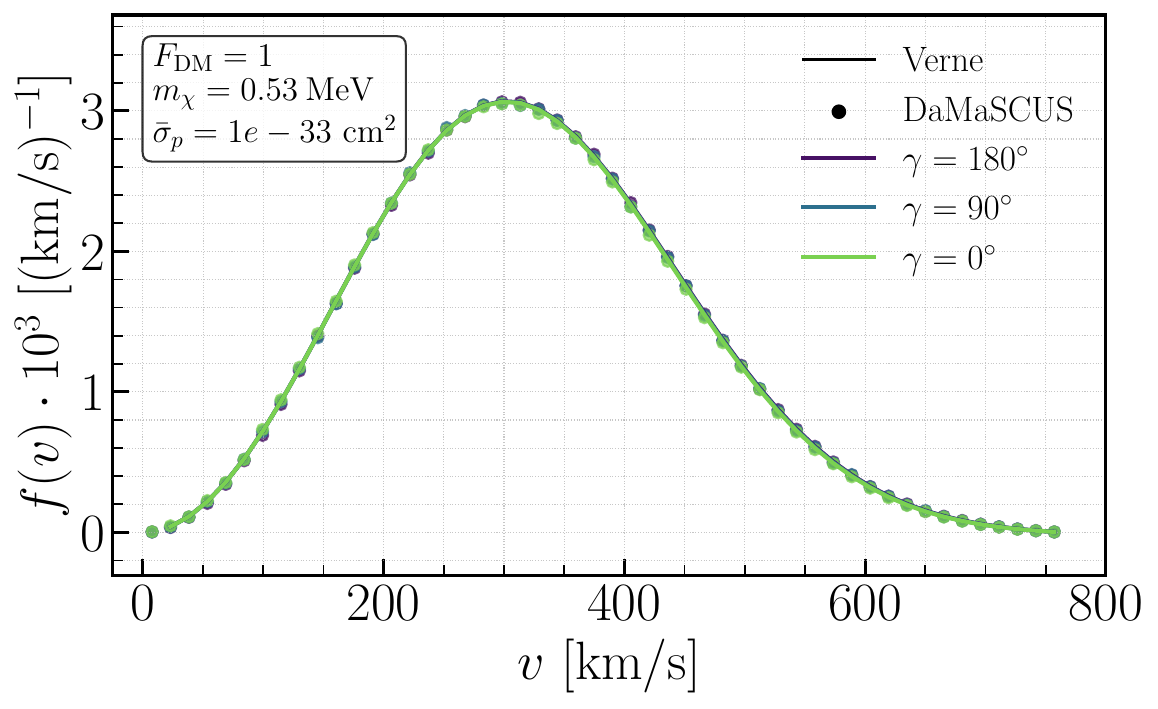}
    \includegraphics[width=0.31\linewidth]{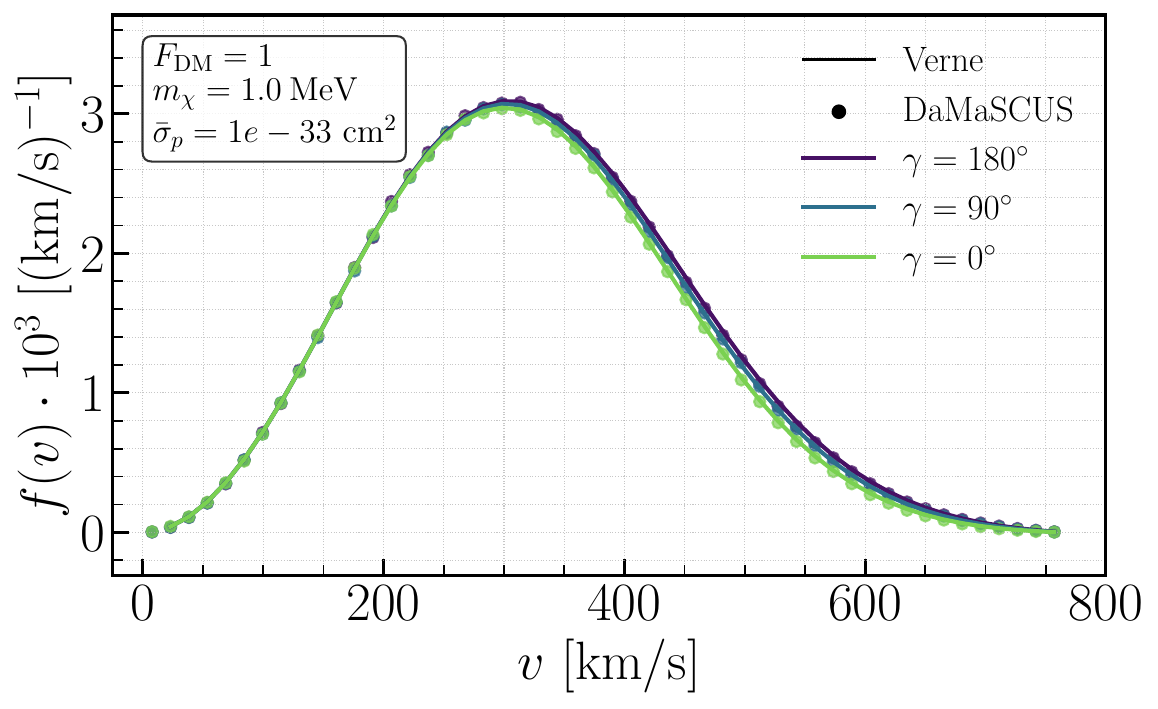}
    \includegraphics[width=0.31\linewidth]{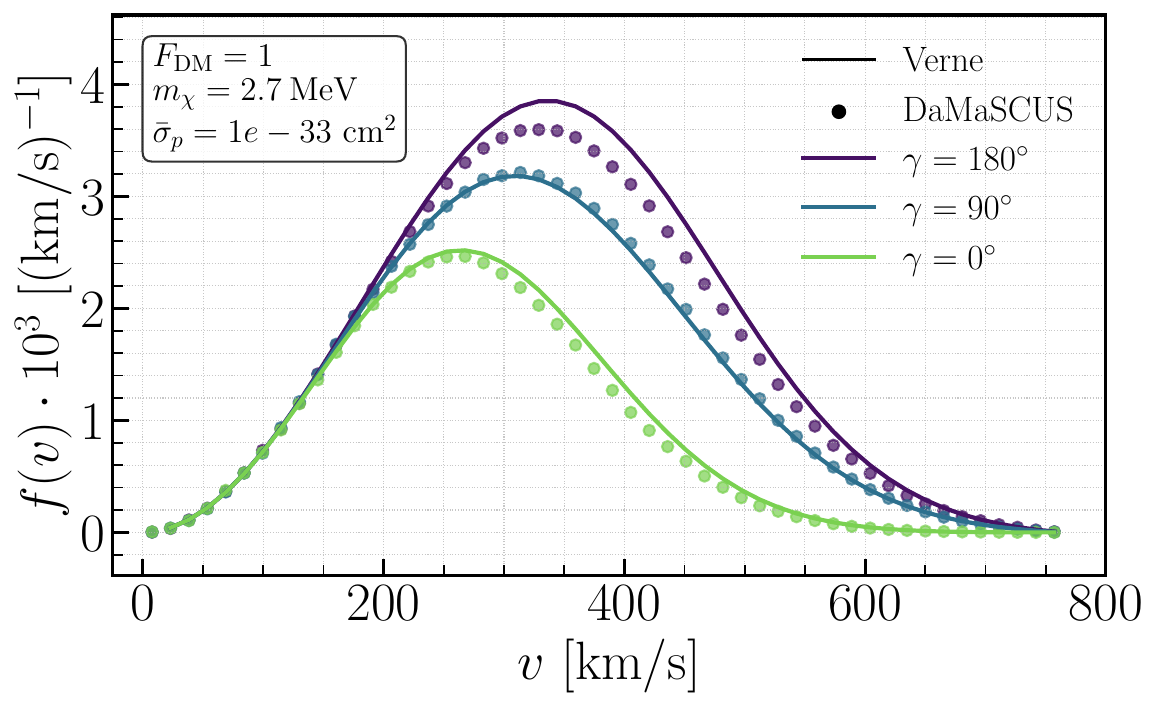}
    \caption{\textbf{Velocity distributions including Earth-scattering in the heavy dark photon mediator model.} Each panel corresponds to a different mass and DM-proton reference cross section. Top-to-bottom, the rows correspond to cross sections of $\bar{\sigma}_p = 10^{-29}\,\mathrm{cm}^2$, $10^{-31}\,\mathrm{cm}^2$, $10^{-33}\,\mathrm{cm}^2$. Left-to-right, the columns correspond to masses of $m_\chi = 0.53\,\mathrm{MeV}$, $m_\chi = 1.0\,\mathrm{MeV}$, $m_\chi = 2.7\,\mathrm{MeV}$. Solid lines show the results obtained using \VerneII while the points are obtained using the \DaMaSCUS Monte Carlo simulations. In each panel we show results for isodetection angles $\gamma = \{0^\circ,\,90^\circ,\, 180^\circ\}$. The former corresponds to the mean DM flux coming from directly below the detector while the latter corresponds to the mean flux coming from directly above.}
    \label{fig:veldists_hm}
\end{figure*}

Figure~\ref{fig:veldists_hm} shows the DM velocity distribution at a detector at a depth of $1400\,\mathrm{m}$ for a subset of the DM masses and cross sections we study here, assuming a heavy dark photon mediator. In each panel, we show the results for three different values of the isodetection angle: $\gamma = 0$ (mean DM flux from directly below the detector), $\gamma = 90^\circ$, and $\gamma = 180^\circ$ (mean DM flux from directly above the detector). Results of the \VerneII calculations presented here are shown as solid lines, with the output from our \DaMaSCUS simulations shown as points. 

Our semi-analytic \VerneII calculations reproduce the general trends seen in the full Monte Carlo simulations. In all cases, the velocity distribution for $\gamma = 0$ shows close agreement between the two approaches. This is the scenario in which the mean DM flux comes from directly below, meaning that the majority of DM particles must cross the entire Earth in order to reach the detector. In this case, the flux arriving at the detector is almost always dominated by the particles which arrive without scattering (the 0-scatter population described in Sec.~\ref{sec:probabilities}), for which \VerneII provides an exact description. 

We note also that at low velocities, the \VerneII and \DaMaSCUS results are in close agreement and in fact match the unscattered `free' DM velocity distribution. This is due to the suppression of the DM-nucleus cross section at low velocities in the heavy mediator case (see Fig.~\ref{fig:cross_section}). At higher velocities, the effects of Earth-scattering become increasing pronounced, with a large suppression for $\gamma = 0$ and an enhancement for $\gamma = 180^\circ$, coming from DM particles which predominantly come from directly overhead and are reflected back in the Earth's bulk after passing the detector.

Compared to \DaMaSCUS, the formalism we have developed here typically under-predicts the velocity distribution for $\gamma = 180^\circ$ for very large cross-sections. For the parameter point $(m_\chi, \,\bar{\sigma}_p) = (0.53\,\mathrm{MeV},\,10^{-29}\,\mathrm{cm}^2)$, shown in the top left panel in Fig.~\ref{fig:veldists_hm}, the \VerneII velocity distribution is around 10-20\% smaller than that produced by $\DaMaSCUS$ at high velocities. This arises because we include only up to 2 scatters in the calculation of the Earth-scattering effect, leading to a reduced reflection probability compared to the full Monte Carlo calculation. However, given that for this parameter point the mean free path of the DM particles in the Earth's crust is $\lambda \sim 0.04\,R_\oplus$, the semi-analytic calculation does well to capture the behaviour at large cross sections. 

Increasing the cross-section further, however, leads to the clear breakdown of the 2-scatter assumption. The parameter point $(m_\chi, \,\bar{\sigma}_p) = (2.7\,\mathrm{MeV},\,10^{-29}\,\mathrm{cm}^2)$, shown in the top right panel in Fig.~\ref{fig:veldists_hm}, gives a mean free path of $\lambda \sim 10^{-5}\,R_\oplus$. In this case, multiple scattering events in the overburden directly above the detector cannot be ignored and including only the $0$- and $2$-scatter contributions suppresses the flux which arrives at the detector from above, in turn suppressing the flux available to be reflected in the bulk of the Earth on the far side, leading to a substantial under-estimate of the velocity distribution for $\gamma = 180^\circ$.

\begin{figure*}[tb]
    \centering
    \includegraphics[width=0.31\linewidth]{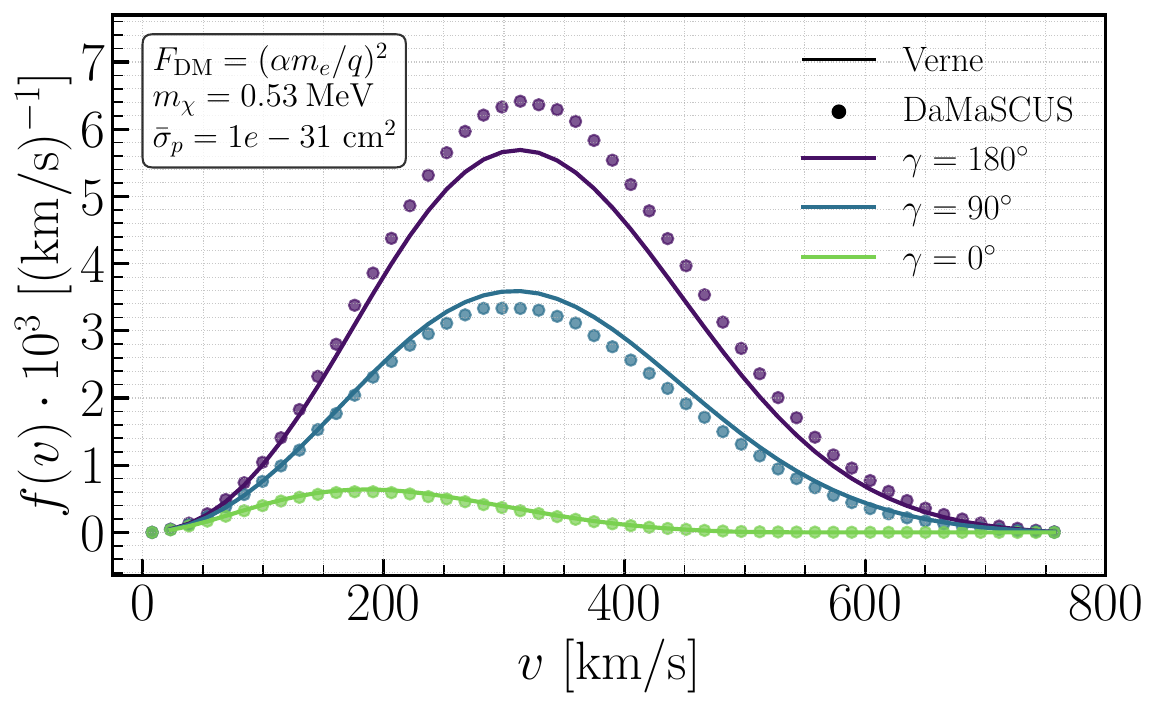}
    \includegraphics[width=0.31\linewidth]{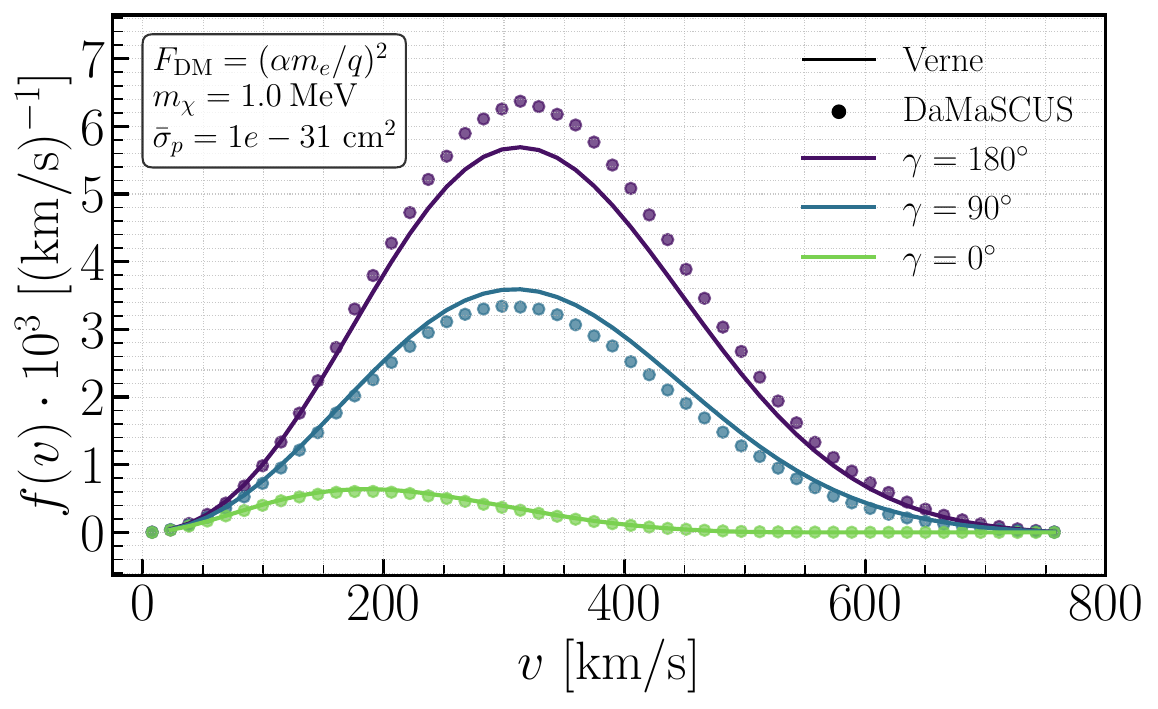}
    \includegraphics[width=0.31\linewidth]{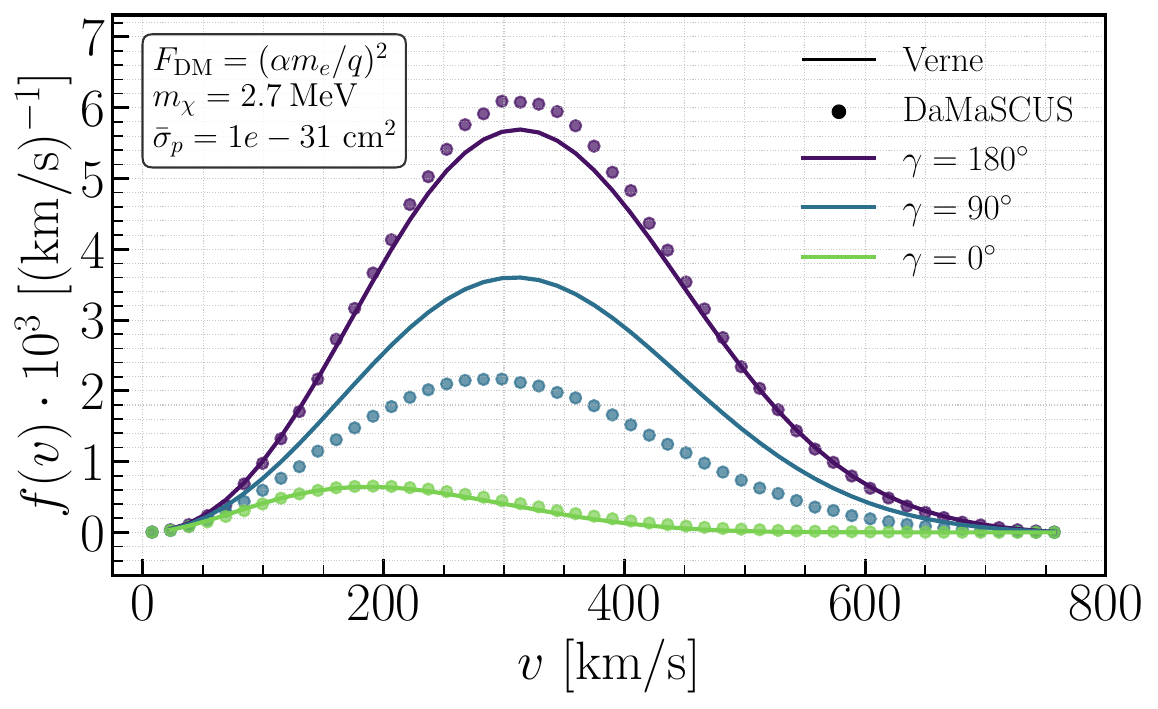}
    \includegraphics[width=0.31\linewidth]{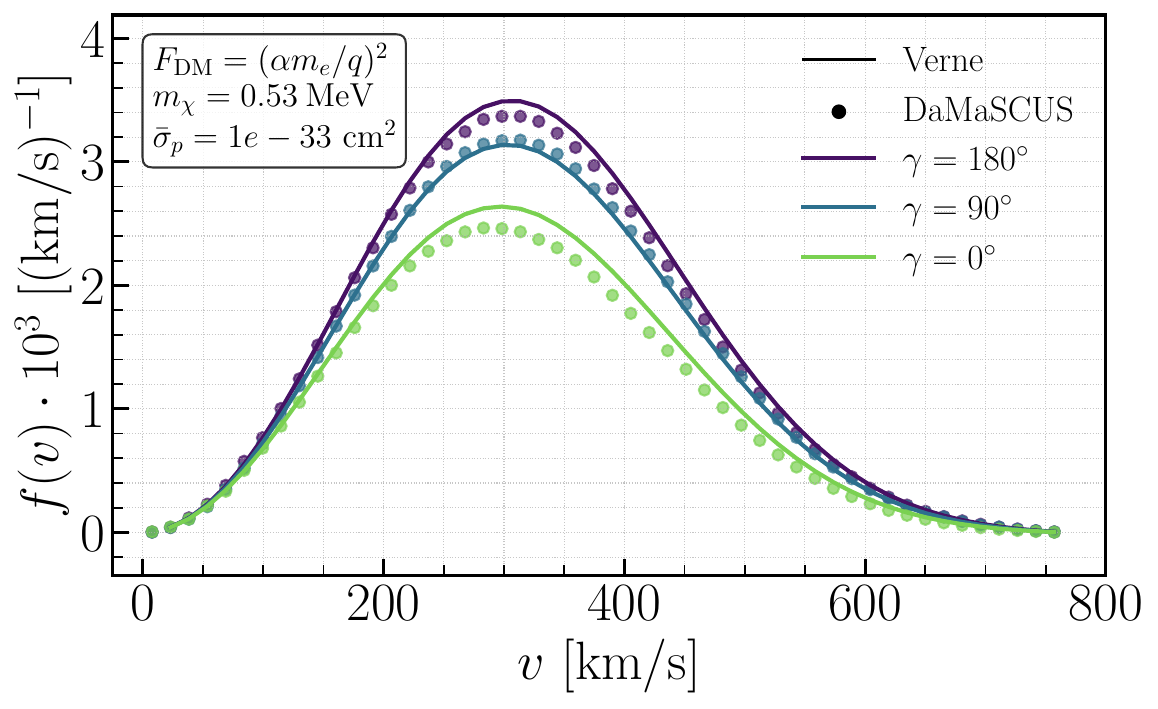}
    \includegraphics[width=0.31\linewidth]{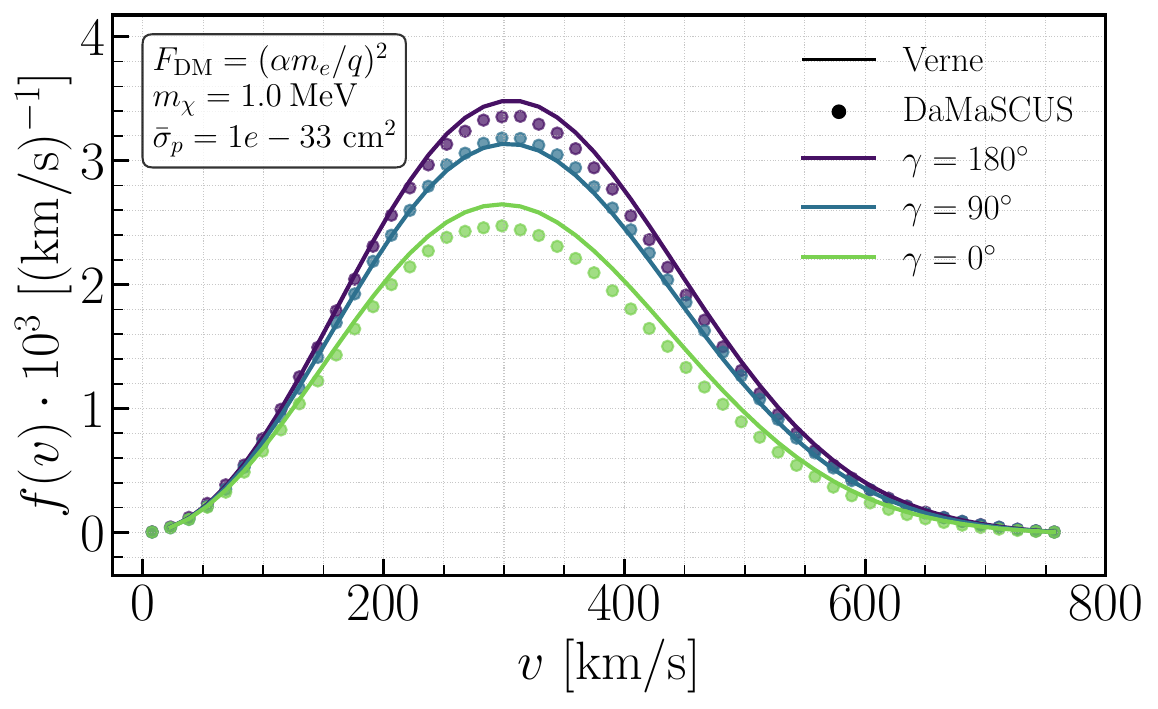}
    \includegraphics[width=0.31\linewidth]{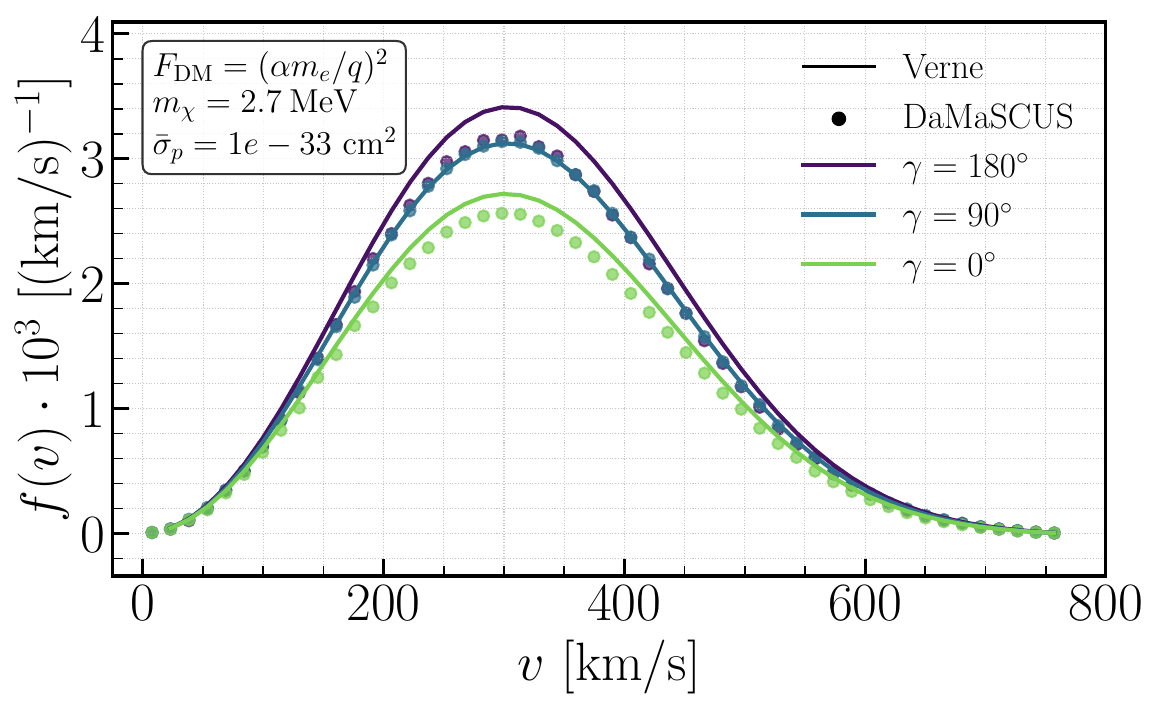}
    \includegraphics[width=0.31\linewidth]{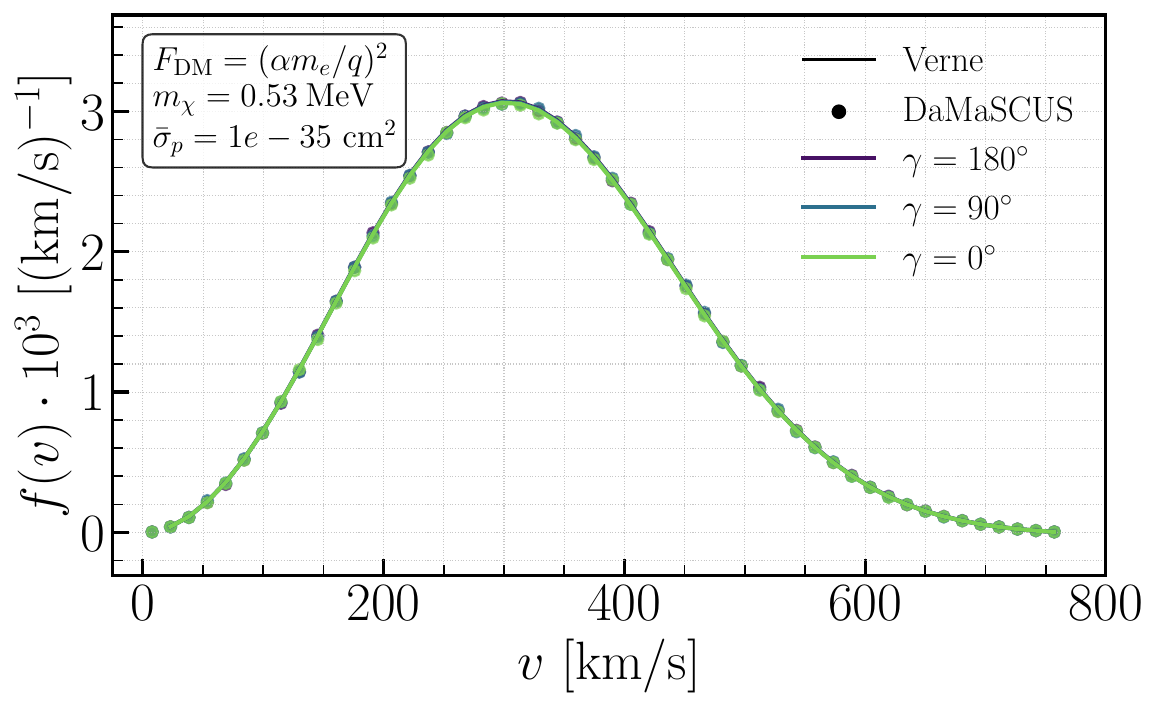}
    \includegraphics[width=0.31\linewidth]{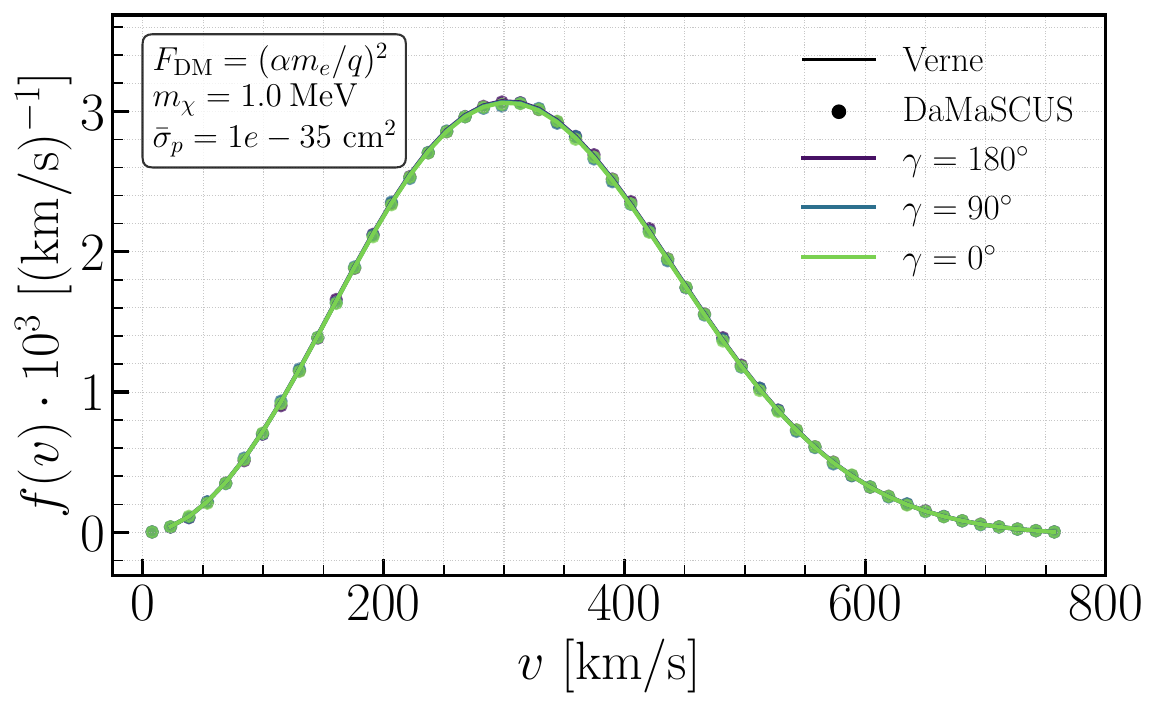}
    \includegraphics[width=0.31\linewidth]{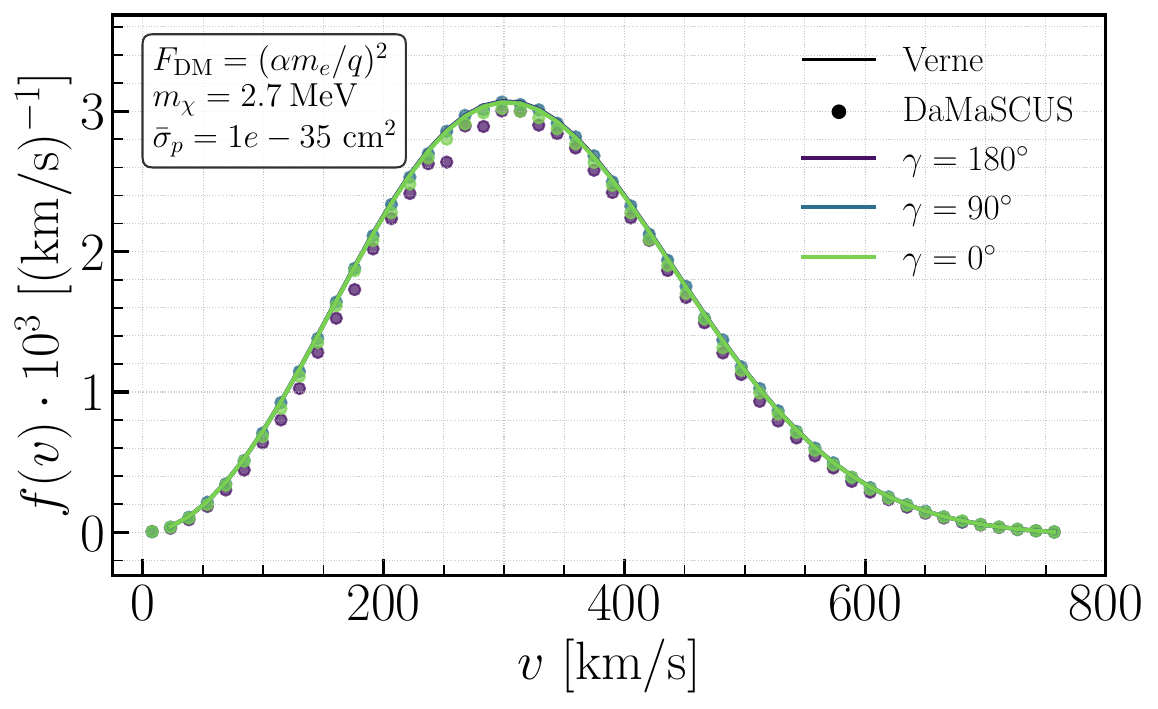}
    \caption{\textbf{Velocity distributions including Earth-scattering in the ultra-light dark photon mediator model.} Top-to-bottom, the rows are $\bar{\sigma}_p = 10^{-31}\,\mathrm{cm}^2$, $10^{-33}\,\mathrm{cm}^2$, $10^{-35}\,\mathrm{cm}^2$. Left-to-right, the columns are $m_\chi = 0.53\,\mathrm{MeV}$, $m_\chi = 1.0\,\mathrm{MeV}$, $m_\chi = 2.7\,\mathrm{MeV}$. See the caption of Fig.~\ref{fig:veldists_hm} for further details.}
    \label{fig:veldists_ulm}
\end{figure*}

Figure~\ref{fig:veldists_ulm} shows some examples of DM velocity distribution in the ultra-light dark photon mediator scenario. Here, we show DM-proton cross sections of $\bar{\sigma}_p = 10^{-31}\,\mathrm{cm}^2$, $10^{-33}\,\mathrm{cm}^2$, $10^{-35}\,\mathrm{cm}^2$ (i.e.\ each row corresponds to a reference cross-section 2 orders of magnitude smaller than the corresponding row in Figure~\ref{fig:veldists_hm}). As in the heavy mediator case, \VerneII typically provides a good estimate of the velocity distribution for the case of $\gamma = 0$ as well as a mild under-estimation of the velocity distribution for $\gamma = 180^\circ$. In contrast, however, we see that for the ultra-light mediator there can be a substantial Earth-scattering effect even at low velocities (see e.g.\ the top left panel of ~\ref{fig:veldists_ulm}). This is because the total DM-nucleus cross section for the ultra-light mediator is not suppressed at low velocity. Instead, the cross section is expected to receive a mild suppression at high velocity (see Fig.~\ref{fig:cross_section}). This in turn means that \VerneII framework (which assumes at most 2 scatters) remains accurate even for larger reference cross sections, leading to good agreement with the Monte Carlo results.

\subsection{DM-electron scattering rates}

While the velocity distributions in the previous section provide an overall picture of the performance of the \VerneII formalism, direct detection experiments do not probe the entire range of DM velocities. Indeed, for low-mass DM, the relevant DM velocities are those in the high-velocity tail; these are the only particles with sufficient kinetic energy to produce an observable signature in the detector. As described in Sec.~\ref{sec:Scattering}, while DM-nucleus scattering typically dominates the Earth-scattering effects for dark photon mediators, we focus here on direct detection via electron scattering, which is most relevant for MeV-scale DM.

For concreteness, we will consider a Silicon semi-conductor target, as used in experiments such as DAMIC-M~\cite{Privitera:2024tpq,DAMIC-M:2024ooa,DAMIC-M:2025luv} and SENSEI~\cite{SENSEI:2023zdf,SENSEI:2024yyt,SENSEI:2025qvp}, which has a band-gap of $\Delta E_\mathrm{Si} = 1.12\,\mathrm{eV}$. For a DM-electron scattering event in which the electron gains an energy $E_e$ with a momentum transfer $q$, the minimum DM velocity required by kinematics is~\cite{Essig:2011nj,Essig:2015cda}:
\begin{equation}
    v_\mathrm{min}(E_e, q) = \frac{E_e}{q} + \frac{q}{2 m_\chi}\,.
\end{equation}
This is minimized for $q = \sqrt{2 m_\chi E_e}$, for which $v_\mathrm{min} = \sqrt{E_e/(2 m_\chi)}$. The mean energy required to produce an electron-hole pair in Silicon is $E_{eh} \approx 3.6\,\mathrm{eV}$~\cite{Ramanathan:2020fwm}, leading to a typical minimum required velocity of
\begin{equation}
    v_\mathrm{min} = \left(\frac{1\,\mathrm{MeV}}{m_\chi}\right)^{1/2}\,400\,\mathrm{km/s}\,.
\end{equation}
Assuming that the semi-conductor can be read out with single-electron resolution, only velocities above this minimum will be relevant to the observable DM-electron scattering rate in Silicon.

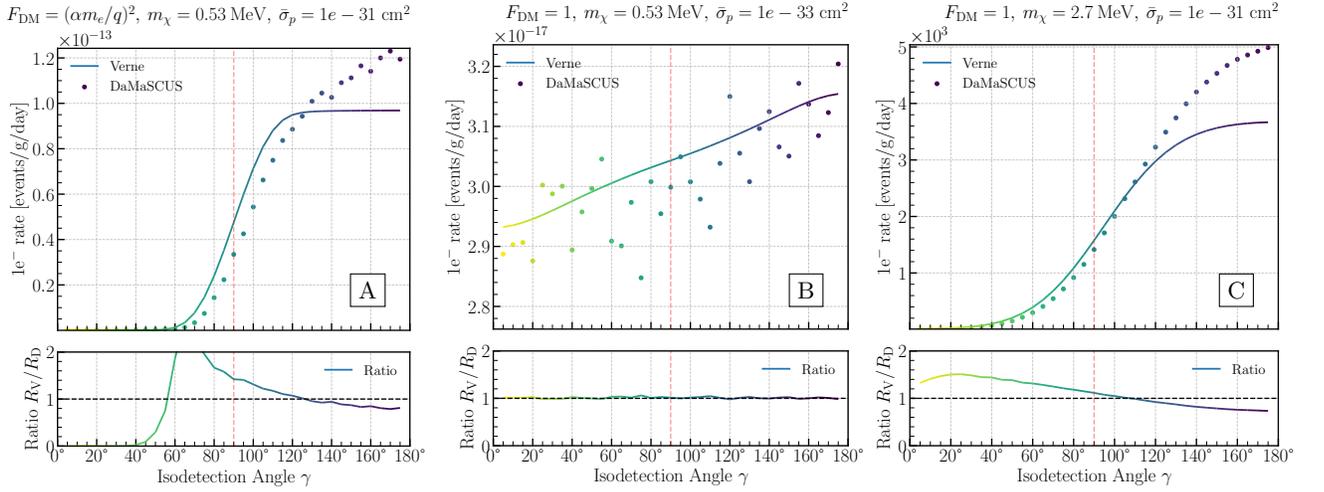
\begin{figure*}[tb]
    \centering
    \trimtikz{ 
        \begin{tikzpicture}
        \node[anchor=south west] (image) at (0,0) {
	       \includegraphics[width=0.32\linewidth]{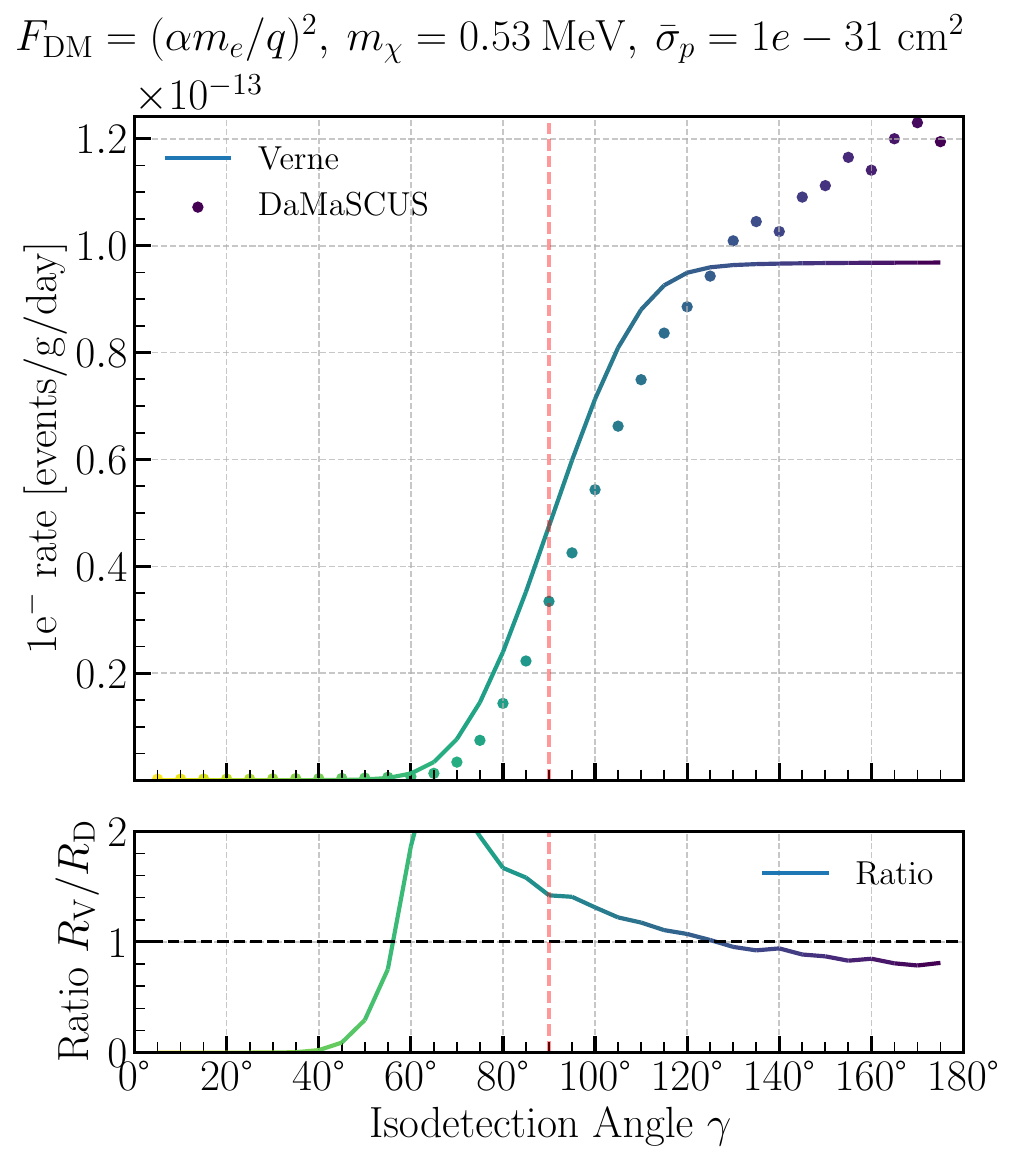}
        };
        \node[draw,align=left,fill=white] at (5,2.8) {A};
        \end{tikzpicture}
        }
        \trimtikz{ \begin{tikzpicture}
        \node[anchor=south west] (image) at (0,0) {
    \includegraphics[width=0.32\linewidth]{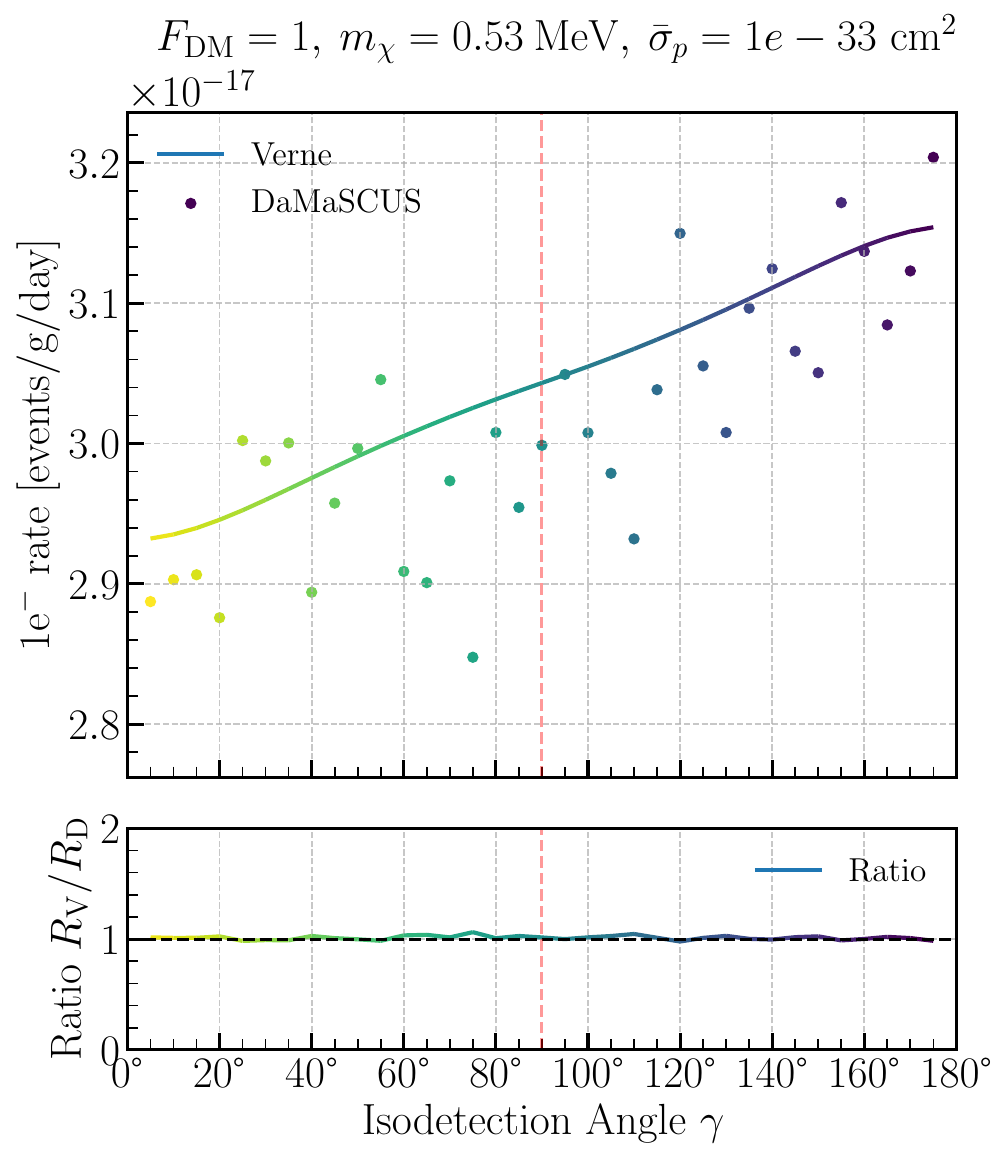}
        };
        \node[draw,align=left,fill=white] at (5,2.8) {B};
    \end{tikzpicture}}
        \trimtikz{ \begin{tikzpicture}
        \node[anchor=south west] (image) at (0,0) {
    \includegraphics[width=0.32\linewidth]{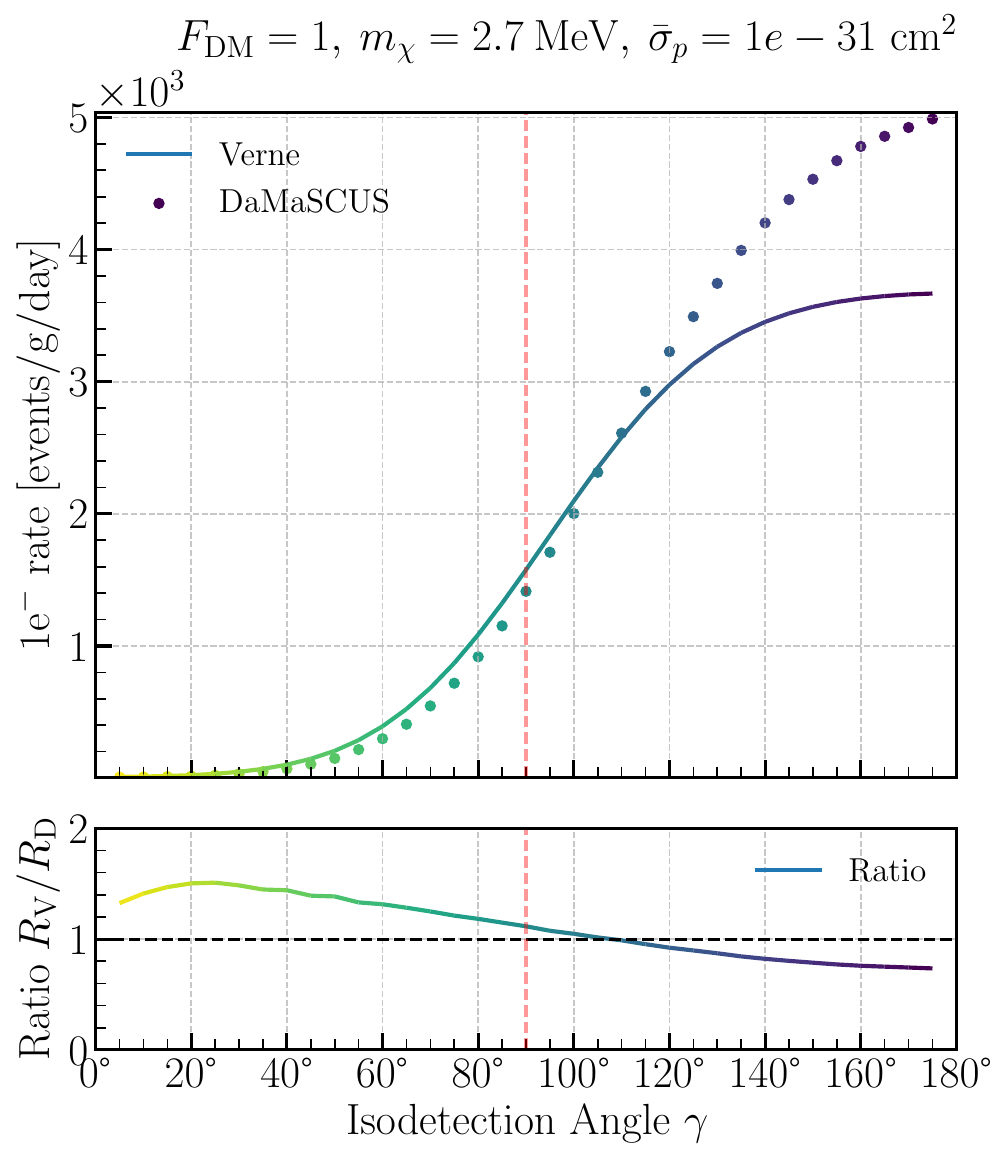}
        };
        \node[draw,align=left,fill=white] at (5,2.8) {C};
    \end{tikzpicture}}
    \caption{\textbf{Comparison of single-electron rates in Silicon as a function of isodetection angle.} In the upper panels, we compare the rate calculated using \textsc{DarkELF}~\cite{Knapen:2021bwg}, with the velocity distributions calculated using \VerneII (solid line) $R_\mathrm{V}$ and \DaMaSCUS $R_\mathrm{D}$ (points). In the lower panel, we show the ratio $R_\mathrm{V}/R_\mathrm{D}$. The DM mass, cross section and mediator model are labeled above each plot. These three example parameter points are labelled A, B, C in Fig.~\ref{fig:FractionalError}. }
    \label{fig:Combined_rates}
\end{figure*}

In order to calculate the DM-electron scattering rates, we use the publicly available \textsc{QEdark} code~\cite{Essig:2015cda} which calculates the DM-induced electronic transition rate as a function of energy, taking into account the band structure of Silicon. More recent codes such as \textsc{DarkELF}~\cite{Knapen:2021bwg} have taken into account in-medium screening effects~\cite{Knapen:2021run}, while codes such as \textsc{EXCEED-DM}~\cite{Trickle:2022fwt}
 and \textsc{QCDark}~\cite{Dreyer:2023ovn} perform more accurate modeling of the electronic band structure. Here, we focus on \textsc{QEdark} for comparison with previous works (e.g.~Ref.~\cite{DAMIC-M:2023hgj}) but of course the \VerneII framework we present can be straightforwardly interfaced with other scattering-rate codes. We map from the energy spectrum of electron recoils to the rate of single-electron ($1e^-$) events using the appropriate ionization probabilities~\cite{Ramanathan:2020fwm}.

In Fig.~\ref{fig:Combined_rates}, we show the single-electron event rates as a function of isodetection angle $\gamma$ for three example parameter points. 
In the left panel of Fig.~\ref{fig:Combined_rates}, we show the event rate for $m_\chi = 0.53\,\mathrm{MeV}$, $\bar{\sigma}_e \approx 2.4 \times 10^{-32}\,\mathrm{cm}^2$ for the ultra-light mediator model. The corresponding DM-proton cross section is $\bar{\sigma}_p = 10^{-31}\,\mathrm{cm}^2$, for which the velocity distributions were shown in the top left panel of Fig.~\ref{fig:veldists_ulm}. The under-estimate of the velocity distribution for $\gamma = 180^\circ$ is directly reflected here in an under-estimate of the signal rate by a factor of at most $\sim 20\%$ for DM fluxes coming directly from above. For $\gamma \sim 90^\circ$, however, \VerneII over-predicts the rate by a factor of $\sim 30\%$. As $\gamma \rightarrow 0^\circ$, the \Verne rate drops roughly to zero, while \DaMaSCUS predicts a non-zero flux, arising from DM particles which are able to cross the Earth and arrive at the detector after under-going a relatively large number of scattering events. This flux is not captured in our semi-analytic formalism, which is limited to at-most 2 scattering events. 

\begin{figure}[tb]
    \centering
    \includegraphics[width=0.99\linewidth]{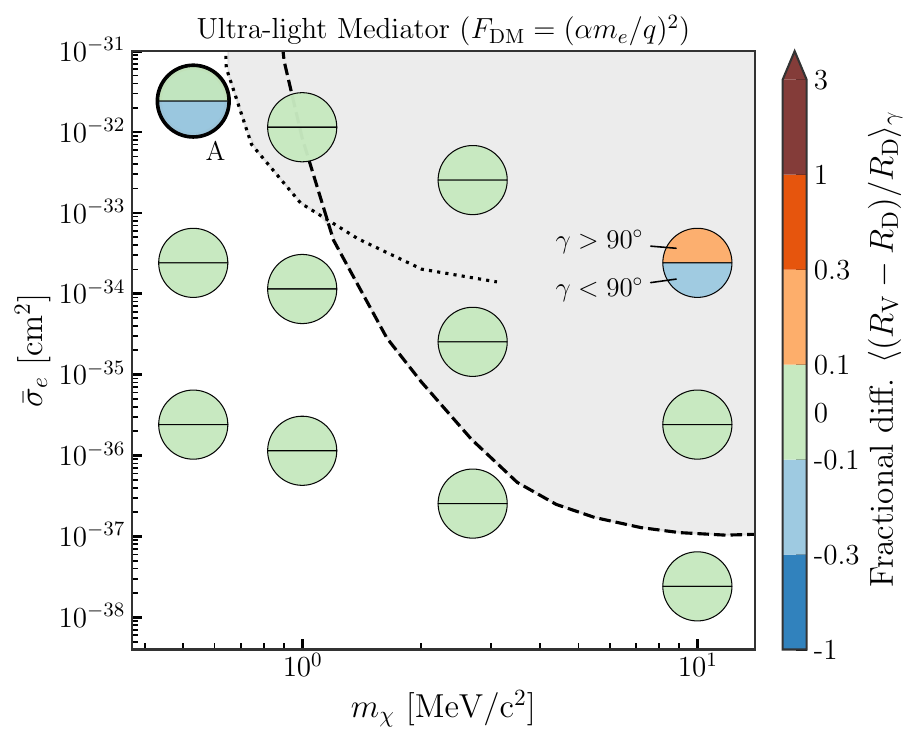}\\
    \includegraphics[width=0.99\linewidth]{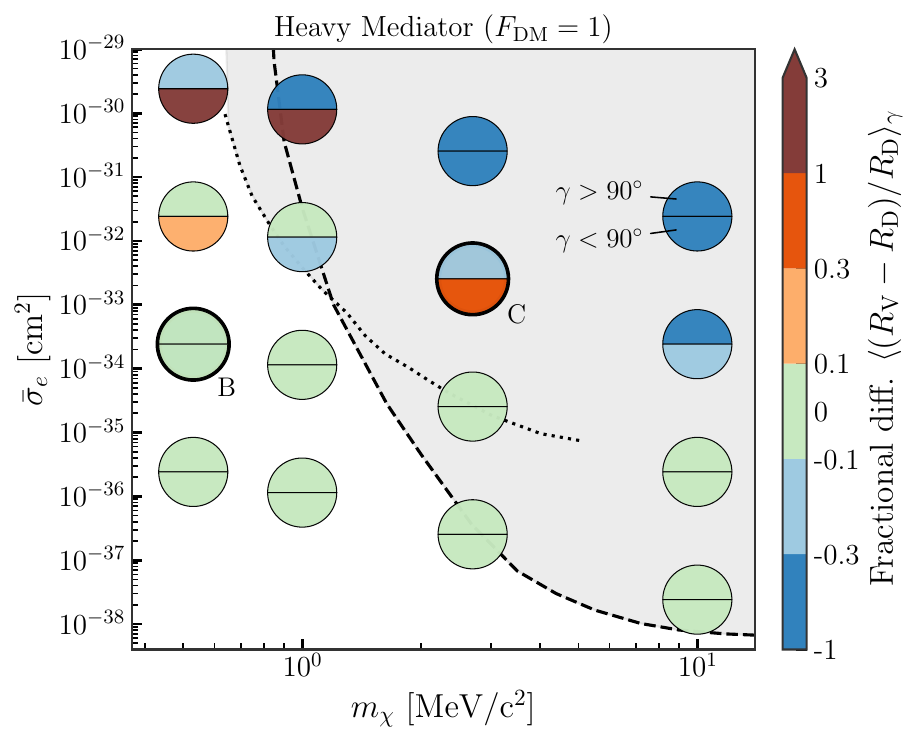}
    \caption{\textbf{Comparison of the DM-electron scattering rate in Silicon between \Verne (V) and \DaMaSCUS (D).} Each point in the parameter space is colored according to the fractional error in the total DM-electron scattering rate $(R_\mathrm{V} - R_\mathrm{D})/R_\mathrm{D}$; darker red (blue) coloring indicates that \Verne overpredicts (underpredicts) the rate compared to \DaMaSCUS. The upper hemisphere of each point corresponds to the rate averaged over isodetection angles $\gamma > 90^\circ$, as would typically be observed for experiments in the Northern hemisphere of the Earth. The lower hemisphere corresponds to $\gamma < 90^\circ$ (Southern Hemisphere). 
    The black dashed line shows the limit from an unmodulated search with DAMIC-M~\cite{DAMIC-M:2025luv} while the black dotted line shows the limit from a modulated search with SENSEI~\cite{SENSEI:2025qvp} (see Ref.~\cite{DAMIC-M:2023hgj} for a similar search with DAMIC-M). Rates for the points labelled A, B and C are shown in Fig.~\ref{fig:Combined_rates}.}
    \label{fig:FractionalError}
\end{figure}

In order to quantify the level of agreement, in Fig.~\ref{fig:FractionalError}, we show the fractional difference in the total rate:
\begin{equation}
\label{eq:fractional_error}
    \frac{\Delta R}{R} \equiv \left\langle \frac{R_\mathrm{V} - R_\mathrm{D}}{R_\mathrm{D}}\right\rangle_{\gamma} = \left\langle \frac{R_\mathrm{V}}{ R_\mathrm{D}}\right\rangle_\gamma - 1\,,
\end{equation}
where $R_\mathrm{V}$ and $R_\mathrm{D}$ are the single electron rates calculated using \Verne and \DaMaSCUS respectively. Each colored circle corresponds to one of the benchmark parameter points considered in this work. The average in Eq.~\eqref{eq:fractional_error} is taken over isodetection angles $\gamma$.  In the upper half of each circle in Fig.~\ref{fig:FractionalError}, we take the average over $\gamma \in [90^\circ,\, 180^\circ]$ (particles mostly from above), which would typically be sampled by detectors in the Northern hemisphere. Instead, in the lower half of each point, we take the average over $\gamma \in [0^\circ,\, 90^\circ]$ (particles mostly from below), typical of the Southern hemisphere. Current limits from DAMIC-M~\cite{DAMIC-M:2025luv} and  SENSEI~\cite{SENSEI:2025qvp} are also shown for comparison.  

The parameter point corresponding to the left panel Fig.~\ref{fig:Combined_rates} is labelled ``A" in the upper panel of Fig.~\ref{fig:FractionalError}. The failure of \Verne to correctly capture the DM flux in the range $\gamma \in [0, 90^\circ]$ leads to an overestimate of the rate by $\mathcal{O}(30\%)$, while for detectors tracing $\gamma \in [0, 90^\circ]$, \VerneII remains accurate at the level of 20\% or better.

In the middle panel of Fig.~\ref{fig:Combined_rates}, we show the rates for the parameters $m_\chi = 0.53\,\mathrm{MeV}$, $\bar{\sigma}_e = 2.4 \times 10^{-34}\,\mathrm{cm}^2$ in the heavy mediator model. This point is labelled ``B" in the lower panel of Fig.~\ref{fig:FractionalError} and the corresponding velocity distributions (for $\bar{\sigma}_p = 10^{-33}\,\mathrm{cm}^2$) were shown in the bottom left panel of Fig.~\ref{fig:veldists_hm}. While no modulation is easily visible by eye in the velocity distribution, the rate displays a modulation at the level of a few percent. In this small-cross-section regime, multiple scattering is unimportant and \VerneII is in excellent agreement with \DaMaSCUS (at the percent level). Despite the large-statistics Monte Carlo simulations performed with \DaMaSCUS, however, the scatter in the calculated rate is comparable to the amplitude of the modulation. Instead, our semi-analytic approach allows us to model even small modulation amplitudes without statistical errors, complementary to Monte Carlo methods.

In the right panel of Fig.~\ref{fig:Combined_rates}, we show the signal rate for $m_\chi = 2.7\,\mathrm{MeV}$, $\bar{\sigma}_e = 2.5 \times 10^{-33}\,\mathrm{cm}^2$ in the heavy mediator model. The corresponding DM-proton cross section is $\bar{\sigma}_p = 10^{-31}\,\mathrm{cm}^2$, for which the velocity distributions are shown in the middle-right panel of Fig.~\ref{fig:veldists_hm}. This parameter point has already been excluded by existing DM-electron scattering searches (see the lower panel of Fig.~\ref{fig:FractionalError}, where this point is labelled ``C"). However, it illustrates the growing discrepancy between \Verne and \DaMaSCUS as we increase the DM mass and cross section. Here, the rate is under-estimated by $\sim 30\%$ at $\gamma = 180^\circ$ and over-estimated by a similar fraction at  $\gamma = 0^\circ$. As we move to the upper right corner of either panel in Fig.~\ref{fig:FractionalError}, the agreement between the two approaches worsens. For the heavy mediator, the \VerneII rate is eventually reduced to zero for very large cross sections as the number of scattering events in the Earth becomes very large.

Focusing instead on the lower-left corner of Fig.~\ref{fig:FractionalError}, we see that for points close to the current bounds (and therefore within reach of near-future modulation searches), the rate calculated by \VerneII is typically accurate at the level of $\sim 10\%$ or better. This is a reflection of the fact that in this region of parameter space, the Earth-scattering effect is dominated by small numbers of scatters before reaching the detector, a regime which is well captured by \Verne.

\section{Conclusions}

In this work, we have presented a semi-analytic formalism for calculating the velocity distribution of Dark Matter (DM) in the lab, taking into account the distortion arising from scattering in the atmosphere and Earth. We focus on models of MeV-mass DM which interacts with the Standard Model via a dark photon mediator. In these scenarios, the DM is too light to be detected in standard direct detection experiments searching for nuclear recoils and instead is constrained via its scattering with electrons. Even so, interactions with nuclei dominate the calculation of Earth-scattering effects. 

Our semi-analytic formalism assumes that the DM particles are sufficiently light that they lose no energy on scattering with nuclei. Instead, they travel in straight lines until scattering, at which point they are reflected back along the incoming trajectory with a given probability. Including up to 2 scatters for each particle, we can then determine the velocity distribution at the detector. This work extends the \Verne formalism developed in Ref.~\cite{Kavanagh:2017cru} for super-heavy DM and alongside this paper we released a new version of the code, \VerneII~\cite{Verne2}.

We compare the resulting velocity distributions and DM-electron scattering rates with results from full Monte Carlo simulations with \DaMaSCUS. These calculations represent the first semi-analytic validation of the \DaMaSCUS code for scattering via dark photon mediators, and show good agreement at low cross sections. Despite the simplifying assumptions we invoke, this agreement persists to intermediate cross sections. Crucially, the \VerneII code provides a computational speed-up of roughly a factor $10^4$, which is important for a systematic exploration of the DM parameter space (especially if uncertainties in the DM velocity distribution are included, as in Ref.~\cite{DAMIC-M:2023hgj}). 

Figure~\ref{fig:FractionalError} provides an estimate of the systematic error induced by this semi-analytic formalism, compared to a more detailed Monte Carlo treatment. In regions of parameter space within reach of current and near-future detectors, the \Verne formalism is accurate at the level of 10\% for detectors in the Northern hemisphere, rising to 30\% in the Southern hemisphere. These results indicate that \Verne is well-suited to calculating daily modulations signals from the Earth-scattering of Dark Matter. 

\acknowledgments

The authors would like to thank the Spanish Agencia Estatal de Investigaci\'on (AEI, MICIU) for their support under the Project \textsc{DMpheno2lab} (PID2022-139494NB-I00) financed by MCIN /AEI /10.13039/501100011033 / FEDER, EU. The authors also thank Xavier Bertou for helpful discussions and contributions to preliminary versions of the \VerneII code.

Finally, we acknowledge the use of the Python scientific computing packages NumPy~\cite{numpy} and SciPy~\cite{scipy}, as well as the graphics environment Matplotlib~\cite{Hunter:2007}.

\bibliography{references}

\end{document}